\documentclass[journal]{IEEEtran}
\usepackage{color}
\usepackage{cite}
\usepackage{amssymb,amsfonts,url}
\usepackage{algorithmic}
\usepackage{textcomp}
\usepackage[rgb]{xcolor}
\usepackage{float}
\usepackage{placeins}
\usepackage{psfrag}
\newcommand{\stz}{\rule{0mm}{2.3ex}}
\newcommand{\stzh}{\rule{0mm}{2.6ex}}
\usepackage{amsmath,graphicx,bm}
\usepackage{color,soul,cite}
\usepackage{MnSymbol}
\usepackage{wasysym}
\usepackage{multirow}
\usepackage{makecell}
\usepackage{xcolor,colortbl,hhline}
\usepackage[super]{nth}
\usepackage{subcaption}
\usepackage[colorlinks=true,linkcolor=black,anchorcolor=black,citecolor=black,urlcolor=blue]{hyperref}
\DeclareCaptionSubType*[alph]{table}
\DeclareCaptionLabelFormat{mystyle}{Table~\bothIfFirst{#1}{ ̃}#2:}
\ifCLASSINFOpdf
\else
\fi
\begin{document}
%
\title{Components Loss for Neural Networks\\ in Mask-Based Speech Enhancement}
%
%
%

\author{Ziyi Xu,
        Samy Elshamy, Ziyue Zhao
        and Tim Fingscheidt,~\IEEEmembership{Senior Member,~IEEE}
\thanks{Z. Xu, S. Elshamy, Z. Zhao, and T. Fingscheidt are with the Institute for Communications Technology, \mbox{Technische} \mbox{Universit{\"a}t} Braunschweig, 38106 Braunschweig, Germany (e-mails: $\left \{ \text{ziyi.xu, s.elshamy, ziyue.zhao, t.fingscheidt} \right \}$@tu-bs.de)}}

\maketitle

\begin{abstract}
Estimating time-frequency domain masks for single-channel speech enhancement using deep learning methods has recently become a popular research field with promising results. In this paper, we propose a novel {\it components loss} (CL) for the training of neural networks for mask-based speech enhancement. During the training process, the proposed CL offers separate control over preservation of the speech component quality, suppression of the residual noise component, and preservation of a naturally sounding residual noise component. We illustrate the potential of the proposed CL by evaluating a standard convolutional neural network (CNN) for mask-based speech enhancement. The new CL obtains a better and more balanced performance in almost all employed instrumental quality metrics over the baseline losses, the latter comprising the conventional mean squared error (MSE) loss and also auditory-related loss functions, such as the perceptual evaluation of speech quality (PESQ) loss and the recently proposed perceptual weighting filter loss. Particularly, applying the CL offers better speech component quality, better overall enhanced speech perceptual quality, as well as a more naturally sounding residual noise. On average, an at least 0.1 points higher PESQ score on the enhanced speech is obtained while also obtaining a higher SNR improvement by more than 0.5\,dB, for {\it seen} noise types. This improvement is stronger for {\it unseen} noise types, where an about 0.2 points higher PESQ score on the enhanced speech is obtained, while also the output SNR is ahead by more than 0.5\,dB. The new proposed CL is easy to implement and code is provided at \href{https://github.com/ifnspaml/Components-Loss}{https://github.com/ifnspaml/Components-Loss}.
\label{sec:abs}
\end{abstract}


\IEEEpeerreviewmaketitle

\section{Introduction}
\IEEEPARstart{S}{peech} enhancement aims at improving the intelligibility and perceived quality of a speech signal that has been degraded, e.g., by additive noise. This task becomes very challenging when only a single-channel microphone mixture signal is available without any knowledge about the individual components. Single-channel speech enhancement has attracted a lot of research attention due to its importance in real-world applications, including telephony, hearing aids devices, and robust speech recognition. Numerous speech enhancement methods were proposed in the past decades. The classical method for single-channel speech enhancement is to estimate a time-frequency (TF) domain mask, or, more specifically, to calculate a spectral weighting rule \cite{Ephraim1984,Ephraim1985,Scalart1996,Lotter2005,fodor2011speech}. To obtain the TF domain coefficients for a spectral weighting rule, the estimation of the noise power, the \textit{a priori} signal-to-noise ratio (SNR) \cite{Ephraim1984,Cohen2005a,Gerkmann2008b,Suhadi2011,Elshamy2015,samy_SNR}, and sometimes also the {\it a posteriori} SNR is required. Finally, the spectral weighting rule is applied to obtain the enhanced speech. Thereby it is still common practice to enhance only the amplitudes and leave the noisy phase untouched. However, the performance of these classical methods degrades significantly in low SNR conditions and also in the presence of non-stationary noise \cite{malah1999tracking}. To mitigate this problem, e.g., a data-driven ideal mask-based approach has been proposed in \cite{fingscheidt2006data,Fingscheidt2008}. Therein, Fingscheidt et al.\ use a simple regression for estimating the coefficients of the spectral weighting rules, which reduces the speech distortion while retaining a high noise attenuation. Interestingly, as with neural networks, this approach already allowed the definition of arbitrary loss functions. Note that Erkelens et al.\ published briefly afterwards on data-driven speech enhancement \cite{erkelens2006general,erkelens2007data}.

In recent years, deep learning methods have been developed and used for weighting rule-based (now widely called mask-based) speech enhancement pushing performance limits even further, also in the presence of non-stationary noise \cite{wang2014training,weninger2014discriminatively,huang2014deep,erdogan2015phase,wang2015deep,williamson2016complex,bao2019new}. The powerful modeling capability of deep learning enables the direct estimation of TF masks without any intermediate steps. Wang et al.\ \cite{wang2014training,wang2018supervised} illustrate that the ideal ratio mask-based approach, in general, performs significantly better than spectral envelope-based methods for supervised speech enhancement. Williamson et al.\ \cite{williamson2016complex} propose to use a complex ratio mask which is estimated from the single-channel mixture to enhance both, the amplitude spectrogram and also the phase of the speech. Different from other methods that directly estimate the TF mask, an approach that predicts the clean speech signal while estimating the TF mask inside the network is proposed in \cite{weninger2014discriminatively,huang2014deep}. Therein the TF mask is applied to the noisy speech amplitude spectrum inside the network in an additional multiplication layer. Thus, the output of the network is already the enhanced speech spectrum, and not a mask which is instead learned implicitly. The authors in \cite{weninger2014discriminatively} demonstrate that the new method outperforms the conventional approach, where the TF mask is the training target and hence learned explicitly. In this paper, we estimate the mask implicitly by using convolutional neural networks (CNNs).

For the training of deep learning architectures for both, mask-based \cite{wang2014training,weninger2014discriminatively,huang2014deep,erdogan2015phase,wang2015deep,williamson2016complex}, and regression-based \cite{du2016regression} speech enhancement, most networks use the mean squared error (MSE) as a loss function. The parameters of the deep learning architectures are then optimized by minimizing the MSE between the inferred results and their corresponding targets. In reality, optimization of the MSE loss in training does not guarantee any perceptual quality of the speech {\it component} and of the residual noise {\it component}, respectively, which leads to limited performance \cite{shivakumar2016perception,liu2017perceptually,zhao2019perceptual,christoph2019learning,martin2018deep,koizumi2018dnn,kolbcek2018monaural,naithani2018deep,zhang2018training,fu2018end}. This effect is even more evident when the level of the noise component is significantly higher than that of the speech component in some regions of the noisy speech spectrum, which explains the bad performance at lower SNR conditions when training with MSE. To minimize the global MSE during training, the network may learn to completely attenuate such TF regions \cite{shivakumar2016perception}, a muting effect that is well-known from error concealment under bad channel SNR conditions \cite{fingscheidt1996error,fingscheidt2001softbit}. This can lead to insufficient quality of the speech component and very unnatural sounding residual noise. To keep more speech component details and to constrain the speech distortion to an acceptable level, Shivakumar et al.\ \cite{shivakumar2016perception} assigned a high penalty against speech component removal in the conventional MSE loss function during training, which results in an improvement in speech quality metrics. A perceptually-weighted loss function that emphasizes important TF regions has recently been proposed in \cite{liu2017perceptually,zhao2019perceptual}, improving speech intelligibility.

A more straightforward direction is to utilize the short-time objective intelligibility (STOI) \cite{taal2010short} and the perceptual evaluation of speech quality (PESQ) \cite{ITU862} metrics as a loss function, which could be used to optimize for speech intelligibility and speech quality, respectively, during training \cite{kolbcek2018monaural,naithani2018deep,fu2018end,zhang2018training,martin2018deep,koizumi2018dnn}. Using STOI as an optimization criterion has been studied in \cite{kolbcek2018monaural,zhang2018training,fu2018end}. Fu et al.\ \cite{fu2018end} proposed a waveform-based utterance enhancement method to optimize the STOI score. They also show that combining STOI with the conventional MSE as an optimization criterion can further increase the speech intelligibility. Using PESQ as an optimization criterion is proposed and studied in \cite{zhang2018training,martin2018deep,koizumi2018dnn}. In \cite{martin2018deep}, the authors have amended the MSE loss by integrating parts of the PESQ metric. This proposed loss achieved a significant gain in speech perceptual quality compared to the conventional MSE loss. Zhang et al.\ \cite{zhang2018training} integrated both STOI and PESQ into the loss function, thereby improving speech separation performance.

However, both, original STOI and PESQ, are non-differentiable functions which cannot be used as an optimization criterion for gradient-based learning directly. A common solution is to use differentiable approximations for STOI or PESQ instead of the original expressions \cite{martin2018deep,koizumi2018dnn,kolbcek2018monaural,fu2018end}. Yet, how to find the best approximated expression is still an open question. In \cite{zhang2018training}, the authors propose a gradient approximation method to estimate the gradients of the original STOI and PESQ metrics. Still, these perceptual loss functions do not offer the flexibility of separate control over noise suppression and preservation of the speech component.

In this paper, we propose a novel so-called {\it components loss} (CL) for deep learning applications in speech enhancement. The newly proposed components loss is inspired by the merit of {\it separately measuring} the performance of speech enhancement systems on the speech component and the residual noise component, which is the so-called white-box approach \cite{Gustafsson1996}, \cite{fingscheidt2007quality}, \cite{yuDSP2011a}, \cite{samy_SNR}. The white-box approach allows to measure the performance of mask-based speech enhancement w.r.t.\ three major aspects: (1) noise attenuation, (2) naturalness of residual noise, and (3) distortion of the speech component. Note that such component-wise quality metrics have also been adopted in ITU-T Rec.\ P.1100 \cite{ITU1100}, P.1110 \cite{ITU1110}, and P.1130 \cite{ITU1130} to evaluate the performance of hands-free systems. We utilize a CNN structure adapted from \cite{zhao2018convolutional} to illustrate the new components loss in the context of speech enhancement. However, the new loss function is not restricted to any specific network topology or application.

Compared to the use of perceptual losses such as PESQ and STOI \cite{martin2018deep,kolbcek2018monaural}, our proposed components loss (CL) is naturally differentiable for gradient-based learning. In practice, the new loss function does not need any additional training material or extensive computational effort compared to other auditory-related loss functions \cite{shivakumar2016perception,liu2017perceptually}, which makes it very easy to implement and also to integrate into existing systems. A further merit is that the new CL not only focuses on offering a strong noise attenuation and a good speech component quality, but also allows for a more natural residual noise, where the trade-off can be controlled directly. Note that highly distorted residual noise can be even more disturbing than the original unattenuated noise signal for human listeners \cite{yuDSP2011a}. To the best of our knowledge, such a loss function has not yet been proposed before.

The rest of the paper is structured as follows: In Section\,\uppercase\expandafter{\romannumeral2} we describe the investigated speech enhancement task and introduce our mathematical notations. The baseline methods used as reference for evaluation are also introduced in this section. Next, we present our proposed components loss function for mask-based speech enhancement in Section\,\uppercase\expandafter{\romannumeral3}. The experimental setup is provided in Section\,\uppercase\expandafter{\romannumeral4}, followed by the results and discussion in Section\,\uppercase\expandafter{\romannumeral5}. Our work is concluded in Section\,\uppercase\expandafter{\romannumeral6}.
\label{sec:intro}
\section{Notations and Baselines}
\subsection{Notations}
We assume an additive single-channel model for the time-domain microphone mixture $y(n)=s(n)+d(n)$ of the clean speech signal $s(n)$ and the added noise signal $d(n)$, with $n$ being the discrete-time sample index. Since mask-based speech enhancement typically operates in the TF domain, we transfer all the signals to the frequency domain by applying a discrete Fourier transform (DFT). Therefore, let $Y_\ell(k)=S_\ell(k)+D_\ell(k)$ be the respective DFT, and $\left |Y_\ell(k)\right |$, $\left |S_\ell(k)\right |$, and $\left |D_\ell(k)\right |$ be their DFT magnitudes, with frame index $\ell\in\mathcal{L}=\left\{1,2,\ldots,L\right\}$ and frequency bin index $k\in\mathcal{K}=\left \{ 0,1,\ldots,K\!-\!1 \right \}$ with $K$ being the DFT size. In this paper, we only estimate the real-valued mask $M_\ell\left (k \right )\in \mathbb{R}$ to enhance the magnitude spectrogram of the noisy speech and use the untouched noisy speech phase for reconstruction, obtaining the predicted enhanced speech spectrum
\begin{equation} \label{clean_speech_est}
\hat{S}_\ell\left (k \right )=Y_\ell(k)\cdot M_\ell\left(k \right ).
\end{equation}
It is then transformed back to the time domain signal $\hat{s}(n)$ with IFFT followed by overlap add (OLA).
\label{sec:2_1}
\subsection{Baseline Network Topology}
\begin{figure}[t!]
	\psfrag{B}[cc][cr]{$Y_\ell(k)$}
	\psfrag{C}[cc][cc]{$\left| Y_\ell(k) \right |$}
	\psfrag{D}[cc][cc]{$M_{\ell}(k)$}
	\psfrag{E}[cc][cc]{NORM}
	\psfrag{F}[cc][cc]{CNN}
	\psfrag{H}[cc][cl]{$\hat{S}_{\ell}\left (k \right )$}
	\psfrag{K}[cc][cr]{Mask-based CNN: baseline and new}
	\centering
	\centerline{\includegraphics[width=0.45\textwidth]{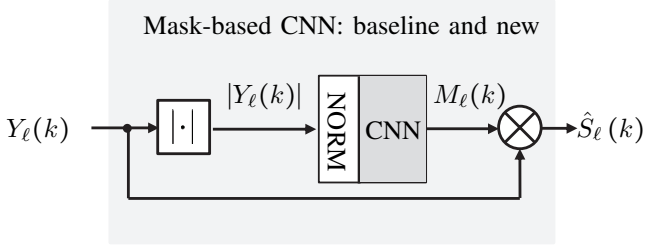}}
	\caption{Schematic of the {\bf mask-based CNN} for speech spectrum enhancement, used for both the baseline CNN (baseline losses) and the new CNN (components loss). Details of the CNN can be seen in Fig.\,\ref{fig:CNN_topology}.}
	\label{fig:system}
\end{figure}

As proposed in \cite{weninger2014discriminatively,huang2014deep}, we predict the clean speech signal while estimating the TF mask inside the network as shown in Fig.\,\ref{fig:system}. The NORM box in Fig.\,\ref{fig:system} represents a zero-mean and unit-variance normalization based on statistics collected on the training set. The CNNs used in this work have exactly the same structure as in \cite[Fig.\,6]{zhao2018convolutional} but with different parameter settings, which will be explained later. This CNN topology has shown great success in coded speech enhancement \cite{zhao2018convolutional}, and is capable of improving speech intelligibility \cite{fu2019metricgan}. Although more complex deep learning architectures could be used, we choose this CNN structure for simple illustration. Note that any other network topology could be used instead.

\begin{figure}[t!]
	\psfrag{a}[cc][cc]{\footnotesize Conv$(F,\!H\!\times\! L_{\rm in})$}
	\psfrag{b}[cc][cc]{\footnotesize Conv$(F,\!H\!\times\! F)$}
	\psfrag{c}[cc][cc]{\footnotesize Maxpooling$(2\!\times\! 1)$}
	\psfrag{d}[cc][cc]{\footnotesize Conv$(2F,\!H\!\times\! F)$}
	\psfrag{e}[cc][cc]{\footnotesize Conv$(2F,\!H\!\times\! 2F)$}
	
	\psfrag{f}[cc][cc]{\footnotesize Maxpooling$(2\!\times\! 1)$}
	\psfrag{g}[cc][cr]{\footnotesize Conv$(F,\!H\!\times\! 2F)$}
	\psfrag{h}[cc][cc]{\footnotesize Upsampling$(2\!\times\! 1)$}
	\psfrag{i}[cc][cc]{\footnotesize Conv$(2F,\!H\!\times\! F)$}
	\psfrag{j}[cc][cc]{\footnotesize Conv$(2F,\!H\!\times\! 2F)$}
	
	\psfrag{k}[cc][cr]{\footnotesize Upsampling$(2\!\times\! 1)$}
	\psfrag{l}[cc][cc]{\footnotesize Conv$(F,\!H\!\times\! 2F)$}
	\psfrag{m}[cc][cc]{\footnotesize Conv$(F,\!H\!\times\! F)$}
	\psfrag{n}[cc][cc]{\footnotesize Conv$(1,\!H\!\times\! F)$}
	
	\psfrag{o}[cc][cc]{\footnotesize $K_{\rm in}\times L_{\rm in}$}
	\psfrag{p}[cc][cc]{\footnotesize $K_{\rm in}\times F$}
	\psfrag{q}[cc][cc]{\footnotesize $K_{\rm in}\times F$}
	\psfrag{r}[cc][cc]{\footnotesize $K_{\rm in}/2\times F$}
	\psfrag{s}[cc][cc]{\footnotesize $K_{\rm in}/2\times 2F$}
	\psfrag{t}[cc][cc]{\footnotesize $K_{\rm in}/2\times 2F$}
	\psfrag{u}[cc][cc]{\footnotesize $K_{\rm in}/4\times 2F$}
	\psfrag{v}[cc][cc]{\footnotesize $K_{\rm in}/4\times F$}
	\psfrag{w}[cc][cc]{\footnotesize $K_{\rm in}/2\times F$}
	\psfrag{x}[cc][cc]{\footnotesize $K_{\rm in}/2\times 2F$}
	\psfrag{y}[cc][cc]{\footnotesize $K_{\rm in}/2\times 2F$}
	\psfrag{z}[cc][cc]{\footnotesize $K_{\rm in}\times 2F$}
	\psfrag{9}[cc][cc]{\footnotesize $K_{\rm in}\times F$}
	\psfrag{8}[cc][rc]{\footnotesize $K_{\rm in}\times F$}
	\psfrag{7}[cc][cc]{\footnotesize $K_{\rm in}\times F$}
	\psfrag{6}[cc][cc]{\footnotesize $K_{\rm in}\times 1$}
	
	\psfrag{A}[cc][cc]{\footnotesize $\mathbf{Y^\prime_\ell}$}
	\psfrag{C}[cc][cc]{\footnotesize $M_\ell(k)$}
	
	\psfrag{5}[cc][cc]{\footnotesize CNN}
	
	\psfrag{4}[cc][cc]{\footnotesize (skip)}
	\psfrag{3}[cc][cc]{\footnotesize \rotatebox{90}{(skip)}}
	\centering
	\centerline{\includegraphics[width=0.45\textwidth]{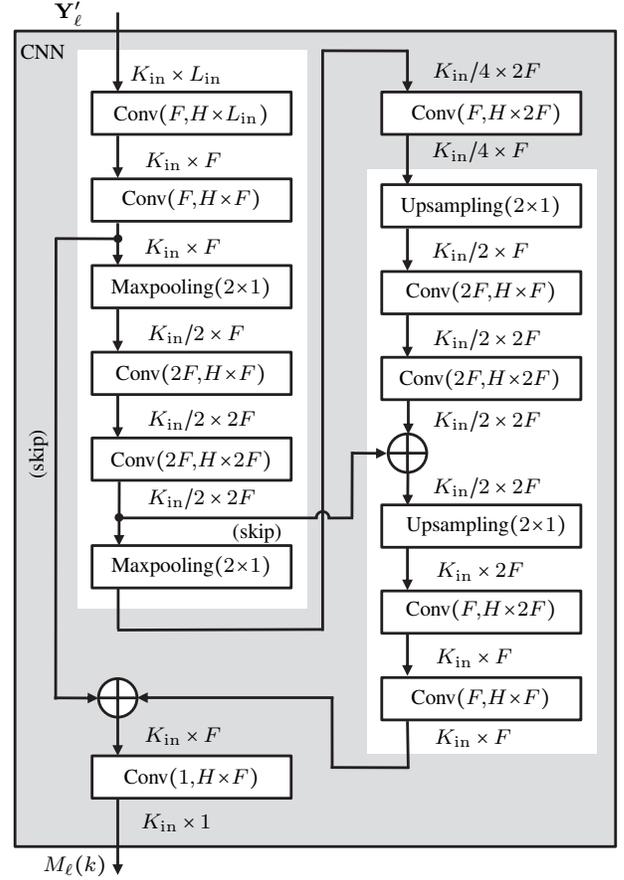}}
	\caption{Topology details of the {\bf employed CNN} in Fig.\,\ref{fig:system} (adopted from \cite[Fig.\,6]{zhao2018convolutional}). The operation Conv$(f,h\times w)$ stands for convolution, with $F$ or $2F$ representing the number of filter kernels in each layer, and $(h\times w)$ represents the kernel size. The maxpooling and upsampling layers have a kernel size of $(2\times 1)$. The stride of maxpooling layers is set to $2$. The gray areas contain two symmetric procedures. All possible forward residual skip connections are added to the layers with matched dimensions.}
	\label{fig:CNN_topology}
\end{figure}

The input of the CNN is a normalized noisy magnitude spectrogram matrix $\mathbf{Y^\prime_\ell}$ with the dimensions $K_{\rm in}\times L_{\rm in}$ as shown in Fig.\,\ref{fig:CNN_topology}, where $K_{\rm in}$ represents the number of input and output frequency bins, and $L_{\rm in}=5$ being the number of normalized context frames centered around the normalized frame $\ell$. Due to the conjugate symmetry of the DFT, it is not necessary to choose $K_{\rm in}$ equal to the DFT size $K$.

The convolutional layers are represented by the \mbox{Conv$(f,h\times w)$} operation in Fig.\,\ref{fig:CNN_topology}. The number of filter kernels is given by $f\in\left \{F, 2F \right \}$ and thus automatically defines also the number of output feature maps which are concatenated horizontally after each convolutional layer. The dimension of the filter kernel is defined by $h\times w$, where $h=H$ is the height and $w\in\left \{L_{\rm in}, F, 2F \right \}$ is the width. The width of the kernel is always corresponding to the width of the respective input to that layer, so that the actual convolution is operating only in vertical (frequency) direction. In the convolution layers, the stride is set to 1, and zero-padding is implemented to guarantee that the first dimension of the layer output is the same as that for the layer input. The maxpooling and upsampling layers have a kernel size of $(2\times 1)$. The stride of the maxpooling layers is set to $2$. The number of the input and output frequency bins $K_{\rm in}$ must be compatible with the two times maxpooling and upsampling operations. All possible forward residual skip connections are added to the layers with matched dimensions to ease any vanishing gradient problems during training \cite{veit2016residual}.

\label{sec:2_2}

\subsection{Baseline Losses}
\label{sec:2_3}
\indent{\bf Baseline MSE}: The conventional approach to train a mask-based CNN for speech enhancement uses the MSE loss. In the training process, the input of the network is the normalized noisy magnitude spectrogram matrix $\mathbf{Y^\prime_\ell}$ as above, and the training target is the corresponding amplitude spectrum of the clean speech $\left |S_\ell(k)\right |$ at frame $\ell$, $k\in\mathcal{K}$. The implicitly estimated mask is applied to the noisy speech amplitude spectrum inside the network as shown in Fig.\,\ref{fig:system}. The MSE loss function for each frame $\ell$ is measured between the clean and the predicted enhanced speech amplitude spectrum, and is defined as
\begin{equation} \label{loss}
J_{\ell}^{\text{MSE}}=\sum_{k\in\mathcal{K}}\left ( \left |\hat{S}_\ell\left (k \right )\right |-\left | S_\ell\left (k \right ) \right | \right )^2.
\end{equation}
As can be observed, all frequency bins have equal importance without any perceptual considerations, such as the masking property of the human ear \cite{zhao2019perceptual}, or the loudness difference \cite{martin2018deep}. Furthermore, as the MSE loss is optimized in a global fashion, the network may learn to completely attenuate some regions of the noisy spectrum, where the noise component is significantly higher compared to the speech component. This behavior can lead to insufficient performance at lower SNR conditions.

{\bf Baseline PW-FILT}: In order to obtain better perceptual quality of the enhanced speech, instead of the MSE loss a so-called perceptual weighting filter loss PW-FILT is used~\cite{zhao2019perceptual}. In this loss, the perceptual weighting filter from code-excited linear prediction (CELP) speech coding is applied to effectively weight the error between the network output and the target. This loss has shown superior performance compared to the MSE loss in speech enhancement~\cite{zhao2019perceptual}, as well as for quantized speech reconstruction~\cite{christoph2019learning}. Some more detail is given in Appendix A.

{\bf Baseline PW-PESQ}: Another option is to adapt PESQ \cite{ITU862}, which is one of the best-known metrics for speech quality evaluation, to be used as a loss function. Since PESQ is a complex and non-differentiable function which cannot be directly used as an optimization criterion for gradient-based learning, a simplified and differentiable approximation of the standard PESQ has been derived and used as a loss function in \cite{martin2018deep}. The proposed PESQ loss is calculated frame-wise from the loudness spectra of the target and the enhanced speech signals. The two distortion terms in the PESQ loss, which consider both auditory masking and threshold effects, are combined with standard MSE to introduce the perceptual criteria. More details are given in Appendix A.

{\bf Baseline PW-STOI}: The maximization of STOI \cite{taal2010short} during training is also the target in several publications \cite{martin2018deep,koizumi2018dnn,kolbcek2018monaural,naithani2018deep,zhang2018training,fu2018end}. In \cite{kolbcek2018monaural}, Kolbcek et al.\ derive a differentiable approximation of STOI, which considers the frequency selectivity of the human ear, for the training of a mask-based speech enhancement DNN. Some more detail is given in Appendix A.

Interestingly, the authors find that no improvement in STOI can be obtained by using the proposed loss function \eqref{PWSTOI_loss}, compared to the conventional network trained using the standard MSE loss function \cite{kolbcek2018monaural}. They conclude in their work that ``the traditional MSE-based speech enhancement networks may be close to optimal from an estimated speech intelligibility perspective" \cite{kolbcek2018monaural}.

Note that PW-STOI is not calculated frame-wise compared to other baseline losses, which makes it very difficult to implement in our setup and to allow a fair comparison. In \cite{kolbcek2018monaural}, the trained network needs to estimate 30 frames of enhanced speech at once, which is represented by $N$ in \eqref{STOI_octave_vector} to calculate the PW-STOI loss. To meet this large output size, the input size can be quite large and unpractical both in our implementation, but due to latency requirements also in practice. Due to the above-cited conclusion from \cite{kolbcek2018monaural} and the large output size requirement, we will not implement PW-STOI loss in our setup.

\section{New Components Loss Functions\\for Mask-Based Speech Enhancement}
The newly proposed components loss (CL) is inspired by the so-called white-box approach \cite{Gustafsson1996}, which utilizes the {\it filtered} clean speech spectrum $\tilde{S}_\ell(k)$ and the {\it filtered} noise component spectrum $\tilde{D}_\ell(k)$ to train the mask-based CNN for speech enhancement as shown in Fig.\,\ref{fig:system2}. We first motivate the use of the white-box approach in the following and then introduce the new components loss.
\subsection{White-Box Approach}
Since our work is inspired by the so-called white-box approach (\hspace{1sp}\cite{Gustafsson1996}, see also \cite{fingscheidt2007quality,yuDSP2011a}), we introduce the {\it filtered} speech spectrum, which is obtained by
\begin{equation} \label{s_tilde}
\tilde{S}_\ell(k)=S_\ell(k)\cdot M_\ell\left(k \right ),
\end{equation}
while the {\it filtered} noise spectrum is estimated by
\begin{equation} \label{n_tilde}
\tilde{D}_\ell(k)=D_\ell(k)\cdot M_\ell\left(k \right ).
\end{equation}
The {\it filtered} speech component spectrum $\tilde{S}_\ell\left (k \right )$ and the {\it filtered} noise component spectrum $\tilde{D}_\ell\left (k \right )$ are transformed back to the time domain signals $\tilde{s}(n)$ and $\tilde{d}(n)$, respectively, with IFFT followed by overlap add (OLA).

Speech enhancement systems aim to provide a strong noise attenuation, a naturally sounding residual noise, and an undistorted speech component. Thus, the evaluation of a speech enhancement algorithm ideally needs to measure the performance w.r.t.\ all three aspects. The white-box approach, which allows to measure the performance based on the {\it filtered} speech component $\tilde{s}(n)$ and the {\it filtered} residual noise component $\tilde{d}(n)$, has been originally proposed in \cite{Gustafsson1996}. A white-box based measure {\it does not} employ the enhanced speech signal $\hat{s}(n)$, but only utilizes the {\it filtered} and {\it unfiltered} components with the {\it unfiltered} ones as a reference \cite{Gustafsson1996,fingscheidt2007quality,yuDSP2011a}. Due to its usefulness, this component-wise white-box measurement has been widely adopted in ITU-T Recs.\ P.1100 \cite{ITU1100}, P.1110 \cite{ITU1110}, and P.1130 \cite{ITU1130} to evaluate the performance of hands-free systems. One might ask whether there is a price to pay with component-wise quality evaluation, since masking effects of human perception are not at all exploited. Accordingly, we will have to use also perceptual quality metrics in the evaluation Sections\,\uppercase\expandafter{\romannumeral4} and \uppercase\expandafter{\romannumeral5}. Interestingly, supporting the adoption of components metrics in ITU-T recommendations, our newly proposed components loss (CL) turns out to be superior both in PESQ and POLQA (perceptual objective listening quality prediction).
\label{sec:3_1}
\subsection{New Components Loss With 2 Components}
\begin{figure}[t!]
	\psfrag{A}[cc][cc]{$\left| S_\ell(k) \right|$}
	\psfrag{C}[cc][cc]{$Y_\ell(k)$}
	\psfrag{E}[cc][cc]{$\left| D_\ell(k) \right|$}
	\psfrag{G}[cc][cc]{$\left| Y_\ell(k) \right |$}
	\psfrag{H}[cc][cc]{$M_{\ell}(k)$}
	\psfrag{I}[cc][cc]{NORM}
	\psfrag{J}[cc][cc]{CNN}
	\psfrag{K}[cc][cc]{New CNN with CL}
	\psfrag{L}[cc][cc]{$\left| \tilde{S}_\ell(k) \right|$}
	\psfrag{P}[cc][cc]{$\left| \tilde{D}_\ell(k) \right|$}
	\centering
	\centerline{\includegraphics[width= 0.45\textwidth]{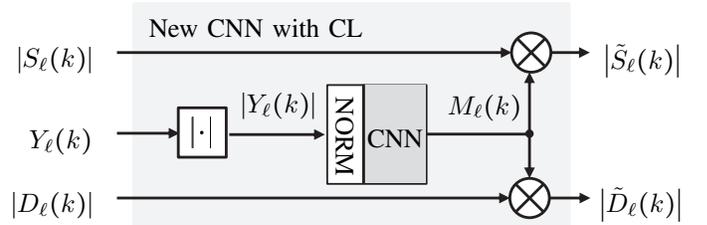}}
	\caption{Proposed {\bf CNN training setup} for speech enhancement according to the {\bf white-box approach}. The hereby applied components loss (CL) is given in \eqref{CLF} and \eqref{CLF2}.}
	\label{fig:system2}
\end{figure}
{\bf New 2CL}: The core innovative step of this work is as follows: Since we assume an additive single-channel model, {\it both}, the amplitude spectrum of the clean speech $\left| S_\ell(k) \right|$, and the additive noise $\left| D_\ell(k) \right|$ are accessible during the training phase, and thus can be used as training targets. First, the {\it filtered} components $\left| \tilde{S}_\ell(k) \right|$ and $\left| \tilde{D}_\ell(k) \right|$ in Fig.\,\ref{fig:system2} are obtained by \eqref{s_tilde} and \eqref{n_tilde}, respectively. Then, we define our proposed components loss (CL) for each frame $\ell$ as
\begin{equation} \label{CLF}
J_{\ell}^{\text{2CL}}=(1-\alpha)\cdot\sum_{k\in\mathcal{K}}\left ( \left |\tilde{S}_\ell\left (k \right )\right |\!-\!\left | S_\ell\left (k \right ) \right | \right )^2+ \alpha\cdot\sum_{k\in\mathcal{K}}\left |\tilde{D}_\ell\left (k \right )\right |^2,
\end{equation}
with $\alpha\in\left [ 0,1 \right ]$ being the weighting factor that can be used to control the trade-off between noise suppression and speech component quality.

This proposed CL \eqref{CLF} dubbed as ``2CL" is the combination of two independent loss contributions, where the first term represents the loss function for the {\it filtered} clean {\it speech} component, and the second term represents the power of the {\it filtered noise} component. Both of the two losses are calculated frame-wise. Minimizing the first term of the loss function is supposed to preserve detailed structures of the speech spectrum, so the perceptual quality of the speech component will be maintained. Any distortion or attenuation being present in the {\it filtered} speech spectrum will be punished by this loss term. The second term of 2CL representing the residual noise power should also be as low as possible. Thus, minimizing the second loss term is responsible for the actual noise attenuation (NA), which is not at all enforced by the first term.

The first and the second term in \eqref{CLF} are combined by the \mbox{weighting} factor $\alpha$. Compared to conventional training using the standard MSE loss function as shown in Fig.\,\ref{fig:system}, our newly proposed training with 2CL offers more information to the network to learn which part of the noisy spectrum belongs to the speech component that should be untouched, and which part is the added noise that should be attenuated. By tuning $\alpha$ close to $1$, 2CL will penalize high residual noise power stronger than severe speech component distortion. Thus, the trained network tends to suppress more noise but maybe at the cost of more speech distortions. When $\alpha$ is close to $0$, the trained network will behave conversely, so that it will offer better speech component quality and may not provide much noise attenuation. Controlling the trade-off between speech component quality and noise attenuation is impossible when using the conventional single-target MSE loss function \eqref{loss}. Note that the enhanced speech $\hat{S}_\ell(k)$ is not part of the loss anymore, only implicitly, keeping in mind that $\hat{S}_\ell(k)=\tilde{S}_\ell(k)+\tilde{D}_\ell(k)$.
\label{sec:3_2}
\subsection{New Components Loss With 3 Components}
{\bf New 3CL}: For a speech enhancement algorithm, a highly distorted residual noise can be even more disturbing than the original unattenuated noise signal for human listeners \cite{yuDSP2011a}. The conventional networks trained with MSE tend to have a strong noise distortion because of the TF bin attenuation behavior as mentioned before. Conversely, the network trained by the proposed 2CL may have less TF bin attenuation, because the TF bin attenuation is also harmful to the speech component and will be penalized by the first term of 2CL. As a consequence, the networks trained by the proposed 2CL are likely to offer more natural residual noise, even though the residual noise {\it quality} is not considered in \eqref{CLF}.

However, to explicitly put the residual noise {\it quality} into consideration during training, we also propose an advanced CL, which is defined as 
\begin{equation} \label{CLF2}
\begin{aligned}
J_{\ell}^{\text{3CL}}&=(1-\alpha-\beta)\cdot\sum_{k\in\mathcal{K}}\left ( \left |\tilde{S}_\ell\left (k \right )\right |\!-\!\left | S_\ell\left (k \right ) \right | \right )^2\\
&+\alpha\cdot\!\sum_{k\in\mathcal{K}}\left |\tilde{D}_\ell\left (k \right )\right |^2\\
&+\beta\cdot\!\sum_{k\in\mathcal{K}}\left(\frac{\left |\tilde{D}_\ell(k)\right |}{\sqrt{\sum_{\kappa\in\mathcal{K}}\left |\tilde{D}_\ell(\kappa)\right |^2}}-\frac{\left |D_\ell(k)\right |}{\sqrt{\sum_{\kappa\in\mathcal{K}}\left |D_\ell(\kappa)\right |^2}}\right )^2,
\end{aligned}
\end{equation}
with $\alpha\in\left [ 0,1 \right ]$ and $\beta\in\left [ 0,1 \right ]$ being the weighting factors to control the speech component quality, the noise suppression, and now also the residual noise quality. In order to have stable training and not to enlarge (!) the speech component MSE (first term in \eqref{CLF2}) during training, we limit the tuning range of the weighting factors to $0\leq\alpha+\beta\leq 1$. This CL with three terms (dubbed ``3CL") is also used to train the speech enhancement neural network as shown in Fig.\,\ref{fig:system2}, without requiring any additional training material compared to when using {2CL}.

The first two terms of the 3CL in \eqref{CLF2} are the same as in \eqref{CLF}, and the additional third term is the loss between the normalized spectra of the {\it filtered} and the {\it unfiltered} noise component, and is supposed to preserve residual noise quality. In order to decouple noise attenuation and residual noise quality, firstly, this additional term is not directly calculated from the {\it filtered} and the {\it unfiltered} noise spectra, but utilizing the {\it normalized} ones. Secondly, both positive and negative differences between the {\it filtered} and the {\it unfiltered} noise spectra are punished equally, which means this loss should be non-negative. So this additional term can have the form of the standard MSE, which is shown in \eqref{CLF2}. This additional loss aims to preserve the residual noise quality even more, enforcing a similarity of residual noise and the original noise component. Note that many alternative definitions of the residual noise quality loss term are possible, however, it should always be ensured that a fullband attenuation ($\tilde{D}_\ell\left (k \right )=\rho\cdot D_\ell(k), \rho<1$) should lead to a zero loss contribution, since it perfectly preserves residual noise {\it quality}.
\label{sec:3_3}

\section{Experimental Setup}
\subsection{Databases and Experimental Setup}
\subsubsection{Database}
The used clean speech data in this work is taken from the Grid corpus \cite{grid_corpus}. The Grid corpus is particularly useful for our experiments, since it provides clean speech samples from many different speakers in a sufficient amount of data for our experiments, which is critical for {\it speaker-independent} training. To make our trained CNN {\it speaker-independent}, we randomly select 16 speakers, containing 8 male and 8 female speakers, and use 200 sentences per speaker for the CNN training. The superimposed noises used in this paper are obtained from the CHiME-3 dataset \cite{chime3}. Both the clean speech and the additive noise signals have a sampling rate of $16\,\text{kHz}$. To generalize the network and also to increase the amount of training data, the noisy speech always contains multiple SNR conditions and includes various noise types. We use pedestrian noise (PED), caf\'e noise (CAFE), and street noise (STR) to generate the training data. We simulate six SNR conditions from $-5\, \text{dB}$ to $20\, \text{dB}$ with a step size of $5\,\text{dB}$. The SNR level is adjusted according to ITU-T P.56 \cite{ITU56}. Thus, the training material consists of $16\times200\times3\times6=57,600$ sentences. From the complete training material, $20\%$ of the data is used for validation and $80\%$ is used for actual training.

During the test phase, the clean speech data is taken from four further Grid speakers, two male and two female, with 10 sentences each neither seen during training nor during validation. The used test noise contains both {\it seen} and {\it unseen} noise {\it types}. The {\it seen} test noise includes PED and CAFE noise, but extracted from different files, which have not been used during training and validation. To perform a noise type-independent test, we additionally create noisy test data using {\it unseen} bus noise (BUS), which is also taken from CHiME-3 and is not seen during training and validation. The test data also contains the six SNR conditions.
\subsubsection{Experimental Setup}
Speech and noise signals are subject to an FFT size of $K=256$, using a periodic Hann window, and $50\%$ overlap. We use the CNN illustrated in Fig.\,\ref{fig:CNN_topology} for the mask estimation. Although more complex deep learning architectures could be used, we choose this CNN structure to illustrate our concept. The number of the input and output frequency bins $K_{\rm in}$ is set to $129+3=132$ for each frame's DFT, as shown in Fig.\,\ref{fig:CNN_topology}. The additional $3$ frequency bins are taken from the redundant bins (from $k=129$ to $k=131$), which are used to make it compatible with the two times maxpooling and upsampling operation in the CNN. The input context is $L_{\rm in}=5$. The number of filters in each convolutional layer represented by $F$ in Fig.\,\ref{fig:CNN_topology} is set to $60$. The used height of the filter kernels is $h=H=15$. In the test phase, we only extract the first $129$ frequency bins from the $132$ output frequency bins to reconstruct the complete spectrum, which is used to obtain the time domain signal by IFFT with OLA. Furthermore, a minibatch size of 128 is used during training. The learning rate is initialized to $2\cdot10^{-4}$ and is halved once the validation loss does not decrease for two epochs. The CNN activation functions are exactly the same as used in \cite{zhao2018convolutional}.

In the baseline training for the perceptual weighting filter loss PW-FILT the linear prediction order represented by $N_{\rm p}$ in \eqref{equ_wgh_filt} is set to $16$. The perceptual weighting factors $\gamma_1$ and $\gamma_2$ in \eqref{equ_wgh_filt} are set to $0.92$ and $0.6$, respectively.
\subsection{Quality Measures}
We use both the white-box approach \cite{Gustafsson1996} which provides the {\it filtered} clean speech component $\tilde{s}(n)$ and the {\it filtered} noise component $\tilde{d}(n)$, as well as standard measures operating on the predicted enhanced speech signal $\hat{s}(n)$. In this paper, we use the following measures \cite{samy_SNR}:\\
\subsubsection{Delta SNR}
$\Delta\text{SNR}=\text{SNR}_{\text{out}}-\text{SNR}_{\text{in}}$, measured in $\text{dB}$. $\text{SNR}_{\text{out}}$ and $\text{SNR}_{\text{in}}$ are the SNR levels of the enhanced speech and the noisy input speech, respectively, and are measured after ITU-T P.56 \cite{ITU56}, based on $\tilde{s}(n)$, $\tilde{d}(n)$ and $s(n)$, $d(n)$, respectively. This measurement should be as high as possible.
\subsubsection{PESQ MOS-LQO}
This measure uses $s(n)$ as reference signal and either the {\it filtered} clean speech component $\tilde{s}(n)$ or the enhanced speech $\hat{s}(n)$ as test signal according to \cite{ITUT_pesq_wb_corri,ITU1110}, being referred to as PESQ$(\tilde{\mathbf{s}})$ and PESQ$(\hat{\mathbf{s}})$, respectively. A high PESQ score indicates better speech (component) perceptual quality.
\subsubsection{Perceptual objective listening quality prediction (POLQA)}
This metric is one of the newest objective metrics for speech quality \cite{ITUT_polqa_2018}. POLQA is measured between the reference signal $s(n)$ and the predicted clean speech $\hat{s}(n)$ according to \cite{ITUT_polqa_2018}, and is denoted as POLQA$(\hat{\mathbf{s}})$. Same as with PESQ, a higher POLQA score is favored.
\subsubsection{Segmental speech-to-speech-distortion ratio}
\[\text{SSDR}=\frac{1}{\left |\mathcal{L}_1 \right |}\sum _{\ell\in\mathcal{L}_1}\text{SSDR}(\ell)\qquad[\text{dB}]\vspace*{-1.5mm} \]
with $\mathcal{L}_1$ denoting the set of speech-active frames \cite{samy_SNR}, and using $\text{SSDR}(\ell)=\max\left \{ \min\left \{ \text{SSDR}'(\ell),30\,\text{dB} \right \}, -10\,\text{dB} \right \}, \text{with}$
$\text{SSDR}'\!(\ell)\!=\!10\log_{10}\!\big(\!\big( \sum\limits_{n\in\mathcal{N}_\ell}\! s(n)^2\big)\!/\! ( \sum\limits_{n\in\mathcal{N}_\ell}\! \left [\! \tilde{s}(n+\Delta)\!-\!s(n)\right ]^2  \big)\!\big),$\vspace*{0.5mm}
with $\mathcal{N}_\ell$ denoting the sample indices $n$ in frame $\ell$, and $\Delta$ being used to perform time alignment of the filtered signal $\tilde{s}(n)$. A low distortion of the filtered speech components leads to a high SSDR. 
\subsubsection{Segmental noise attenuation ($\text{NA}_{\text{seg}}$)}
\begin{equation} \label{NA}
\text{NA}_{\text{seg}}=10\log_{10}\left [ \frac{1}{\left |\mathcal{L}\right |}\sum_{\ell\in\mathcal{L}}\text{NA}_{\text{frame}}(\ell) \right ],\qquad[\text{dB}]
\end{equation}
with
\[\text{NA}_{\text{frame}}(\ell)=\frac{\begin{matrix} \sum_{n\in\mathcal{N}_\ell} d^2(n) \end{matrix}}{\begin{matrix} \sum_{n\in\mathcal{N}_\ell} \tilde{d}^2(n+\Delta) \end{matrix}}.\] We measure $\text{NA}_{\text{seg}}$ for the purpose of parameter optimization, so we can easily choose the weighting factors that offer a strong noise attenuation as well as a good speech component perceptual quality. In the test phase, we use the $\Delta\text{SNR}$ metric to reflect the overall SNR improvement caused by noise suppression instead of using a single $\text{NA}_{\text{seg}}$ metric.
\subsubsection{The weighted log-average kurtosis ratio (WLAKR)}
This metric measures the noise distortion (especially penalizing musical tones) using $d(n)$ as reference signal and the {\it filtered} noise component $\tilde{d}(n)$ as test signal according to ITU-T P.1130 \cite{ITU1130}. A WLAKR score that is closer to zero indicates less noise distortion \cite{yuDSP2011a,yu2012black}. Accordingly, in our analysis we will show averaged {\it absolute} WLAKR values.
\subsubsection{STOI}
We use STOI to measure the intelligibility of the enhanced speech, which has a value between zero and one \cite{taal2010short}. A STOI score close to one indicates high intelligibility.

We group these measurements to noise {\it component} measures ($\Delta$SNR and WLAKR), speech {\it component} measures (SSDR and PESQ$(\tilde{\mathbf{s}})$), and total performance measures (PESQ$(\hat{\mathbf{s}})$, POLQA$(\hat{\mathbf{s}})$, and STOI).
\section{Results and Discussion}
\subsection{Hyperparameter Optimization}
To allow for an efficient hyperparameter search, we optimize the weighting factors for our proposed components loss functions by using only $12.5\%$ of the validation set data. The total performance measures PESQ$(\hat{\mathbf{s}})$, POLQA$(\hat{\mathbf{s}})$, and STOI are averaged over all training noise types and all SNR conditions.
\subsubsection{2CL hyperparameter $\alpha$}
The performance for different weighting factors $\alpha$ for 2CL \eqref{CLF} is shown in Table\,\ref{develop}. The baseline MSE in Table\,\ref{develop} represents the conventional mask-based CNN as shown in Fig.\,\ref{fig:system} and is trained using the MSE loss function. It becomes obvious that a choice of $\alpha$ in \eqref{CLF} being far away from $0.5$ leads to either bad perceptual speech quality or low speech intelligibility. This behavior is expected, since speech enhancement requires a sufficiently strong noise attenuation as well as an almost untouched speech component. To choose the best weighting factor $\alpha$ from Table\,\ref{develop}, we first discard all columns where at least one measure is below or equals the baseline MSE and subsequently select from the remaining values $\alpha\in\left \{0.45, 0.5, 0.55\right \}$ the best performing, which is $\alpha=0.5$. The selected setting is \setlength{\fboxsep}{0pt}\colorbox{black!15}{grey-shaded} as shown in Table\,\ref{develop}.

In Fig.\,\ref{fig:plot2}, we plot the obtained $\text{NA}_{\text{seg}}$ vs. PESQ$(\tilde{\mathbf{s}})$ values for the various combinations of hyperparameters as shown in Table\,\ref{develop}. Here, from top to bottom, each marker depicts a certain SNR condition varying from $20\, \text{dB}$ to $-5\, \text{dB}$ in steps of $5\, \text{dB}$. The further a curve is to the right {\it and} to the top, the better the overall performance. We can see that the performance for the selected hyperparameter $\alpha=0.5$ (dot-dashed pink line, circle markers) is a quite balanced choice.
\begin{table}[t!]
	\caption{Optimization of hyperparameter $\alpha$ for the {\bf new 2CL} \eqref{CLF} on {\bf $\mathbf{12.5\%}$ of the validation set}. The selected setting is \setlength{\fboxsep}{0pt}\colorbox{black!15}{grey-shaded}.}
	\scriptsize
	\centering
	\setlength\tabcolsep{2.5pt}
	\begin{tabular}{|c|c|cccccccc|}
		\hline
		\multirow{2}{*}{} & Baseline & \multicolumn{8}{c|}{\stzh{New $J_{\ell}^{\text{2CL}}(\alpha)$}} \\ \hhline{~~--------} 
		&                     MSE      &  $\alpha=$ &     \stz{0}  &   0.1   &    0.45  &    \cellcolor{black!15}{0.5}  &    0.55  &   0.75 & 0.9  \\ \hhline{----------}
		PESQ$(\hat{\mathbf{s}})$              &  \stz{2.21}   &   & 1.78   &  2.10     &  2.48   &   \cellcolor{black!15}{2.50}   &   2.41   &  2.60 & 2.59   \\ \hhline{----------}
		POLQA$(\hat{\mathbf{s}})$              &  \stz{1.91}   &   & 2.11   &  1.86     &  2.21   &   \cellcolor{black!15}{2.23}   &   2.22   &  2.39 & 2.26   \\ \hhline{----------}
		\stz{STOI}              &               0.72                &    &0.72   &    0.74   &  0.73    &   \cellcolor{black!15}{0.73}   &   0.73   & 0.70 & 0.68   \\ \hhline{----------}
	\end{tabular}
	\label{develop}
	\vspace*{0.34 cm}
\end{table}
\begin{figure}[t!]
	\psfrag{aaaaa}[cc][cc]{\scriptsize $\alpha=0$}
	\psfrag{bbbbbbb}[cc][cc]{\scriptsize $\alpha=0.1$}
	\psfrag{ccccccccc}[cc][cc]{\scriptsize $\alpha=0.45$}
	\psfrag{ddddddd}[cc][cc]{\scriptsize {\setlength{\fboxsep}{0pt}\colorbox{black!15}{$\alpha=0.5$}}}
	\psfrag{eeeeeeee}[cc][cc]{\scriptsize $\alpha=0.55$}
	\psfrag{ffffffffffffffff}[cc][cc]{\scriptsize $\alpha=0.75$}
	\psfrag{ggggggg}[cc][cc]{\scriptsize $\alpha=0.9$}
	\psfrag{h}[cc][cc]{PESQ($\tilde{\mathbf{s}}$)}
	\psfrag{g}[cc][cc]{$\text{NA}_{\text{seg}}$ [dB]}
	\psfrag{U}[cc][cc]{\scriptsize $20$ dB}
	\psfrag{B}[cc][cc]{\scriptsize $-5$ dB}
	\centering
	\centerline{\includegraphics[width=0.45\textwidth]{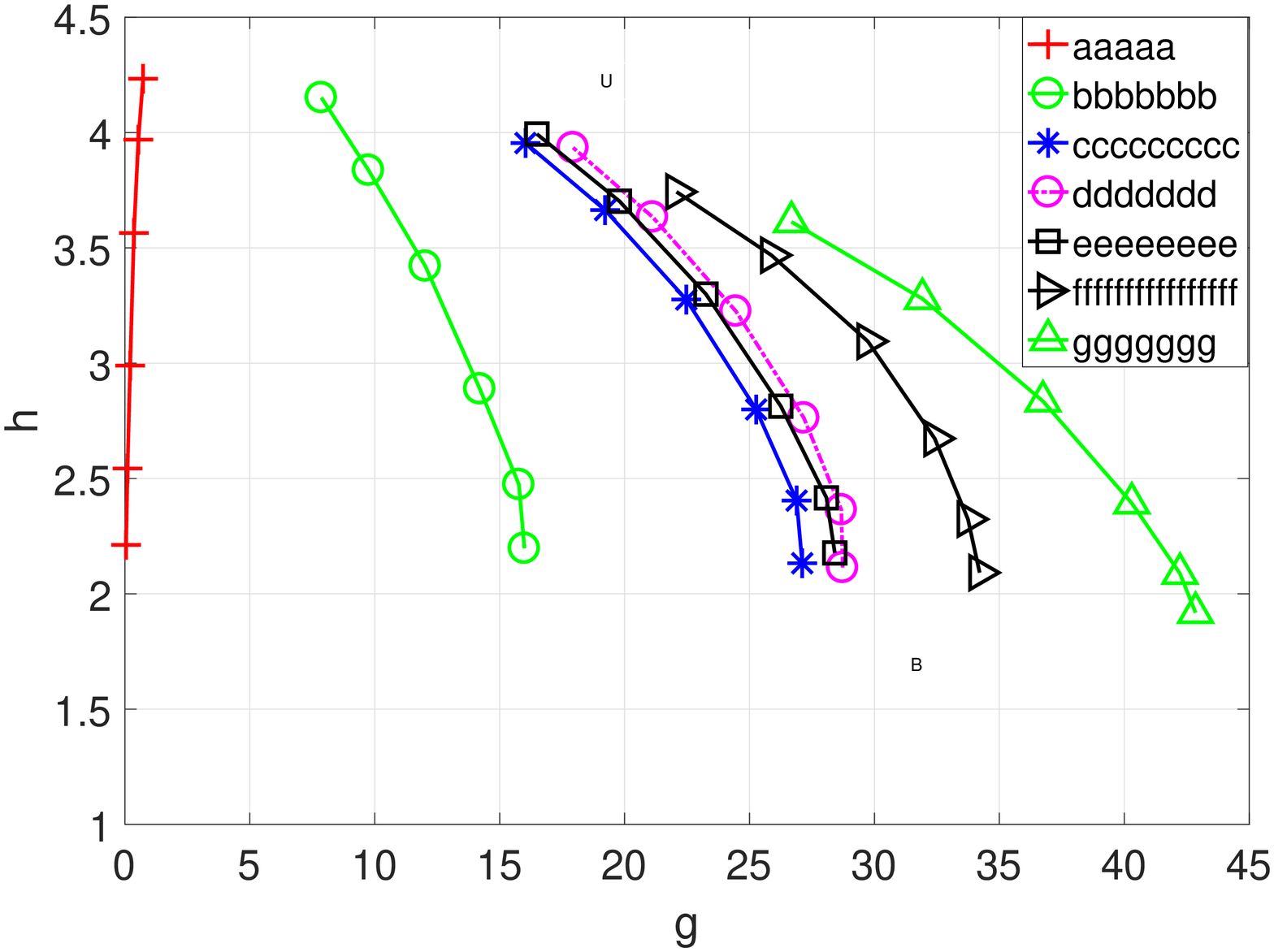}}
	\caption{Noise attenuation $(\text{NA}_{\text{seg}})$ vs. speech {\it component} quality $(\text{PESQ}(\tilde{\mathbf{s}}))$ for different parameters $\alpha$ for the {\bf new 2CL} \eqref{CLF} on {\bf $\mathbf{12.5\%}$ of the validation set}. From top to bottom, the markers are corresponding to six SNR conditions from $20\, \text{dB}$ to $-5\, \text{dB}$ with a step size of $5\,\text{dB}$. The selected setting $\alpha=0.5$ is \setlength{\fboxsep}{0pt}\colorbox{black!15}{grey-shaded} in the legend.}
	\label{fig:plot2}	
\end{figure}
\subsubsection{3CL hyperparameters $\alpha$, $\beta$}
We also optimize the combination of the weighting factors $\alpha$ and $\beta$ for 3CL in \eqref{CLF2} as shown in Table\,\ref{develop2}. The baseline MSE in Table\,\ref{develop2} is the same as the one in Table\,\ref{develop}. Interestingly, a good performance is achieved mostly\footnote{Note that the case of $\alpha=0.6$ and $\beta=0.2$ is also quite good on PESQ and POLQA, but performs poorly on STOI.} when the weighting factors for speech component quality $(1\!-\!\alpha\!-\!\beta)$ and noise attenuation $(\alpha)$ are equal or very close to each other --- as is the case for our 2CL choice of $\alpha=0.5$ in Table\,\ref{develop}. This is the case for 3CL when $\alpha\in\left\{0.05, 0.1, 0.15, 0.2, 0.3, 0.4\right\}$ and the corresponding (in that order) $\beta\in\left\{0.9, 0.8, 0.7, 0.6, 0.4, 0.2\right\}$, as shown in Table\,\ref{develop2} marked by {\Large \textcolor{green}{$\bm *$}}. Thus, tuning the weighting factors for speech component quality and noise attenuation in an unbalanced way will degrade the overall performance, especially for STOI or PESQ as shown in Table\,\ref{develop2}. As the best combination in Table\,\ref{develop2} we select $\alpha=0.1$ and $\beta=0.8$, highlighted by a \setlength{\fboxsep}{0pt}\colorbox{black!15}{grey-shaded} font. The additional term of 3CL \eqref{CLF2}, weighted with $\beta$, is supposed to preserve the residual noise quality. It can further improve the overall performance of PESQ and POLQA as can be seen when comparing the grey-shaded columns of Tables\,\ref{develop} and \ref{develop2}. The reason could be that PESQ and POLQA measures favor natural residual noise.
\begin{table}[t!]
	\caption{Optimization of hyperparameters $\alpha$ and $\beta$ for the {\bf new 3CL} \eqref{CLF2} on {\bf $\mathbf{12.5\%}$ of the validation set}. The selected setting is \setlength{\fboxsep}{0pt}\colorbox{black!15}{grey-shaded}.}
	\scriptsize
	\centering
	\setlength\tabcolsep{1.3pt}
	\begin{tabular}{|c|c|ccccccccccc|}
		\hline
		\multirow{3}{*}{} & Baseline & \multicolumn{11}{c|}{\stzh{New $J_{\ell}^{\text{3CL}}(\alpha, \beta)$}} \\ \hhline{~~-----------} 
		&            MSE      &    \stzh{$\alpha=$} & $0.05^{\textcolor{green}{\bm *}}$  &   {0.1}   &    0.1 &    \cellcolor{black!15}{$0.1^{\textcolor{green}{\bm *}}$}  &   $0.15^{\textcolor{green}{\bm *}}$& $0.2^{\textcolor{green}{\bm *}}$  &  $0.3^{\textcolor{green}{\bm *}}$& $0.4^{\textcolor{green}{\bm *}}$&0.6&0.8    \\ \hhline{~~-----------}
		&                  &     \stzh{$\beta =$}&   $0.9^{\textcolor{green}{\bm *}}$        &    {0.4}    &        0.6   &        \cellcolor{black!15}{$0.8^{\textcolor{green}{\bm *}}$}    &$0.7^{\textcolor{green}{\bm *}}$    &    $0.6^{\textcolor{green}{\bm *}}$  &$0.4^{\textcolor{green}{\bm *}}$    &   $0.2^{\textcolor{green}{\bm *}}$&0.2&0.1\\ \hhline{-------------}
		PESQ$(\hat{\mathbf{s}})$              &  \stz{2.21}   &   & 2.47   &  {2.19}     &  2.24   &   \cellcolor{black!15}{2.54}   & 2.50 & 2.49 & 2.47 &  2.51  &2.59&2.60 \\ \hhline{-------------}
		POLQA$(\hat{\mathbf{s}})$              &  \stz{1.91}   &   & 2.25   &  {1.88}     &  1.97   &   \cellcolor{black!15}{2.28}   &  2.23 &2.23 & 2.20 &  2.26  &2.34&2.24  \\ \hhline{-------------}
		\stz{STOI}              &       0.72                &    &0.73   &    {0.74}   &  0.74    &   \cellcolor{black!15}{0.73}   & 0.73  &0.73 &0.73  & 0.73  &0.71&0.68   \\ \hhline{-------------}
	\end{tabular}
	\label{develop2}
\end{table}
\begin{figure}[t!]
	\psfrag{aaaaaaaaaaaaaa}[cc][cc]{\scriptsize $\alpha=0.05, \beta=0.9$}
	\psfrag{bbbbbbbbbbbbbb}[cc][cc]{\scriptsize {{$\alpha=0.1, \beta=0.4$}}}
	\psfrag{ccccccccccccccccc}[cc][cc]{\scriptsize $\alpha=0.1, \beta=0.6$}
	\psfrag{bbbbbbbbbbbbb1}[cc][cc]{\scriptsize {\setlength{\fboxsep}{0pt}\colorbox{black!15}{$\alpha=0.1, \beta=0.8$}}}
	\psfrag{dddddddddddddd}[cc][cc]{\scriptsize $\alpha=0.15, \beta=0.7$}
	\psfrag{bbbbbbbbbbbbb2}[cc][cc]{\scriptsize $\alpha=0.2, \beta=0.6$}
	\psfrag{eeeeeeeeeeeeee}[cc][cc]{\scriptsize $\alpha=0.3, \beta=0.4$}
	\psfrag{fffffffffffffffffffffffffff}[cc][cc]{\scriptsize $\alpha=0.4, \beta=0.2$}
	\psfrag{ggggggggggggg}[cc][cc]{\scriptsize $\alpha=0.6, \beta=0.2$}
	\psfrag{hhhhhhhhhhhhhhhh}[cc][cc]{\scriptsize $\alpha=0.8, \beta=0.1$}
	\psfrag{h}[cc][cc]{PESQ($\tilde{\mathbf{s}}$)}
	\psfrag{g}[cc][cc]{$\text{NA}_{\text{seg}}$ [dB]}
	\psfrag{U}[cc][cc]{\scriptsize $20$ dB}
	\psfrag{B1}[cc][cc]{\scriptsize $-5$ dB}
	\psfrag{B2}[cc][cc]{\scriptsize $-5$ dB}
	\psfrag{S}[cc][cc]{\Large \textcolor{green}{$\bm *$}}
	\centering
	\centerline{\includegraphics[width=0.45\textwidth]{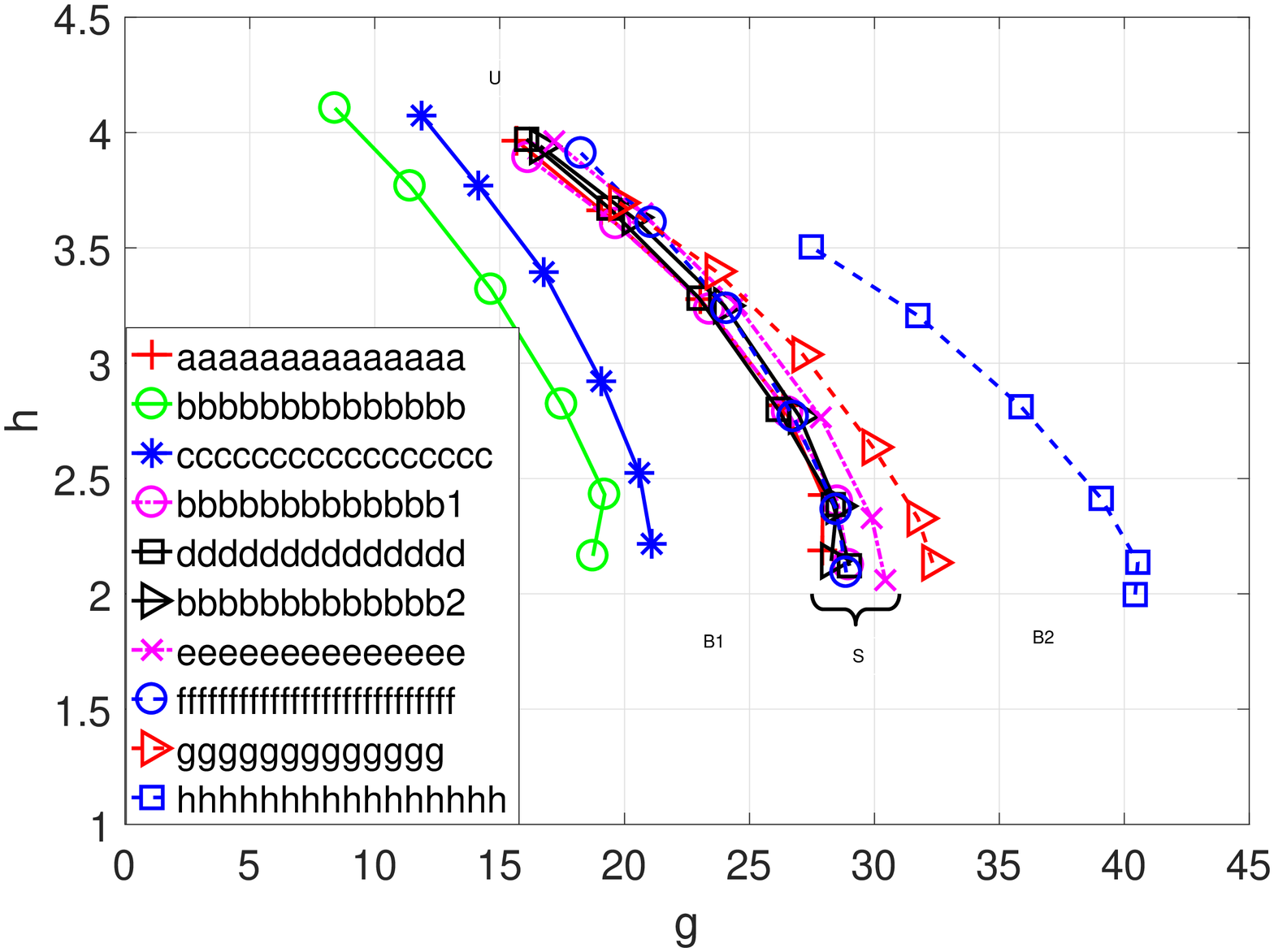}}
	
	\caption{Noise attenuation $(\text{NA}_{\text{seg}})$ vs. speech {\it component} quality $(\text{PESQ}(\tilde{\mathbf{s}}))$ for different parameters $\alpha$ and $\beta$ for the {\bf new 3CL} \eqref{CLF2} on {\bf $\mathbf{12.5\%}$ of the validation set}. From top to bottom, the markers are corresponding to six SNR conditions from $20\, \text{dB}$ to $-5\, \text{dB}$ with a step size of $5\,\text{dB}$. The selected setting is \setlength{\fboxsep}{0pt}\colorbox{black!15}{grey-shaded} in the legend. All curves marked by {\Large \textcolor{green}{$\bm *$}} fulfill $\alpha=1\!-\!\alpha\!-\!\beta$, meaning that the noise attenuation and the speech distortion contribute equally to the 3CL loss \eqref{CLF2}.}
	\label{fig:plot}
\end{figure}

For the combinations of hyperparameters in Table\,\ref{develop2}, we also plot $\text{NA}_{\text{seg}}$ vs. PESQ$(\tilde{\mathbf{s}})$ as shown in Fig.\,\ref{fig:plot}. All curves marked by {\Large \textcolor{green}{$\bm *$}} fulfill $\alpha=1\!-\!\alpha\!-\!\beta$, meaning that the noise attenuation and the speech distortion contribute equally to the 3CL loss \eqref{CLF2}. Obviously, these curves show a comparably good speech component quality as well as a strong noise attenuation at the same time. The overall difference between these curves are very small, which is also reflected in Table\,\ref{develop2}. In Fig.\,\ref{fig:plot}, the curve for $\alpha=0.8$ and $\beta=0.1$ shows very strong noise attenuation, but with quite low PESQ$(\tilde{\mathbf{s}})$. This is expected since the contribution of the noise attenuation in 3CL loss \eqref{CLF2}, which is controlled by $\alpha$, is the strongest from the investigated values. On the contrary, when $\alpha=0.1$ and the corresponding $\beta\in\left\{0.4, 0.6\right\}$, we obtain the highest PESQ$(\tilde{\mathbf{s}})$ and the weakest noise attenuation. Our selected hyperparameter combination (dot-dashed pink line, circle markers) is among the curves marked by {\Large \textcolor{green}{$\bm *$}} showing quite balanced performance.
\subsubsection{PW-PESQ hyperparameters $\lambda_1$, $\lambda_2$}
\begin{table}[t!]
	\caption{Optimization of hyperparameters $\lambda_1$ and $\lambda_2$ for {\bf baseline PW-PESQ} \eqref{PWPESQ_loss} on {\bf $\mathbf{12.5\%}$ of the validation set}. The selected setting is \setlength{\fboxsep}{0pt}\colorbox{black!15}{grey-shaded}.}
	\scriptsize
	\centering
	\setlength\tabcolsep{2.5pt}
	\begin{tabular}{|c|c|ccccccccc|}
		\hline
		\multirow{3}{*}{} & Baseline & \multicolumn{9}{c|}{\stzh{Baseline $J_{\ell}^{\text{PW-PESQ}}(\lambda_1, \lambda_2)$}} \\ \hhline{~~---------} 
		&            MSE      &    $\lambda_1 =$& \stz{0.01}  &   0.1   &    \cellcolor{black!15}{0.2} &    0.4  &   0.5& 0.6  &  0.8& 0.9    \\ \hhline{~~---------}
		&                  &     $\lambda_2 =$&   \stz{0.99}        &    0.9    &        \cellcolor{black!15}{0.8}   &        0.6    &0.5    &    0.4  &0.2    &   0.1\\ \hhline{-----------}
		PESQ$(\hat{\mathbf{s}})$              &  \stz{2.21}   &   & 2.22   &  2.22     &  \cellcolor{black!15}{2.23}   &   2.18   & 2.18 & 2.15 & 2.12 &  2.21    \\ \hhline{-----------}
		POLQA$(\hat{\mathbf{s}})$              &  \stz{1.91}   &   & 1.90   &  1.87     &  \cellcolor{black!15}{1.89}   &   1.84   & 1.87 & 1.81 & 1.84 &  1.85    \\ \hhline{-----------}
		\stz{STOI}              &       0.72                &    &0.71   &    0.72   &  \cellcolor{black!15}{0.72}    &   0.73   & 0.73  &0.72 &0.72  & 0.72    \\ \hhline{-----------}
	\end{tabular}
	\label{develop3}
\end{table}
\begin{figure}[t!]
	\psfrag{aaaaaaaaaaaaaaaaaaa}[cc][cc]{\scriptsize $\lambda_1=0.01, \lambda_2=0.99$}
	\psfrag{bbbbbbbbbbbbbbbbbbb}[cc][cc]{\scriptsize $\lambda_1=0.1, \lambda_2=0.9$}
	\psfrag{cccccccccccccccccccccc}[cc][cc]{\scriptsize {\setlength{\fboxsep}{0pt}\colorbox{black!15}{$\lambda_1=0.2, \lambda_2=0.8$}}}
	\psfrag{bbbbbbbbbbbbbbbbb1}[cc][cc]{\scriptsize $\lambda_1=0.4, \lambda_2=0.6$}
	\psfrag{dddddddddddddddddd}[cc][cc]{\scriptsize $\lambda_1=0.5, \lambda_2=0.5$}
	\psfrag{bbbbbbbbbbbbbbbbb2}[cc][cc]{\scriptsize $\lambda_1=0.6, \lambda_2=0.4$}
	\psfrag{eeeeeeeeeeeeeeeeeee}[cc][cc]{\scriptsize $\lambda_1=0.8, \lambda_2=0.2$}
	\psfrag{fffffffffffffffffffffffffffffffff}[cc][cc]{\scriptsize $\lambda_1=0.9, \lambda_2=0.1$}
	\psfrag{h}[cc][cc]{PESQ($\tilde{\mathbf{s}}$)}
	\psfrag{g}[cc][cc]{$\text{NA}_{\text{seg}}$ in dB}
	\psfrag{U}[cc][cc]{\scriptsize $20$ dB}
	\psfrag{B}[cc][cc]{\scriptsize $-5$ dB}
	\centering
	\centerline{\includegraphics[width=0.45\textwidth]{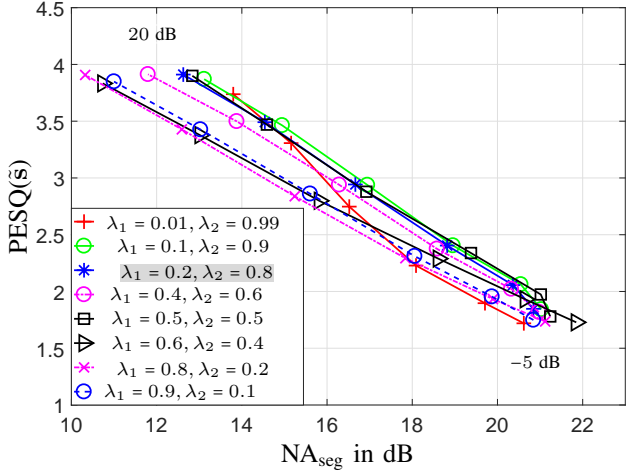}}
	
	\caption{Noise attenuation $(\text{NA}_{\text{seg}})$ vs. speech {\it component} quality $(\text{PESQ}(\tilde{\mathbf{s}}))$ for different parameters $\lambda_1$ and $\lambda_2$ for the {\bf baseline PW-PESQ} \eqref{PWPESQ_loss} on {\bf $\mathbf{12.5\%}$ of the validation set}. From top to bottom, the markers are corresponding to six SNR conditions from $20\, \text{dB}$ to $5\, \text{dB}$ with a step size of $-5\,\text{dB}$. The selected setting is \setlength{\fboxsep}{0pt}\colorbox{black!15}{grey-shaded} in the legend.}
	\label{fig:plot3}
\end{figure}
For the baseline loss function $J_{\ell}^{\text{PW-PESQ}}$, to allow a fair comparison, the weighting factors $\lambda_1$ and $\lambda_2$ in \eqref{PWPESQ_loss} are also optimized, and the results are shown in Table\,\ref{develop3}. To limit the range of tuning parameters, we define $\lambda_1+\lambda_2\leqslant 1$. Since optimizing $J_{\ell}^{\text{PW-PESQ}}$ during training aims to improve the perceptual quality of the enhanced speech, we choose the optimal weighting factors, with which the best PESQ$(\hat{\mathbf{s}})$ is achieved. Furthermore, we discard the settings that offer a STOI lower than the baseline MSE. The selected setting $\lambda_1=0.2$, $\lambda_2=0.8$ in Table\,\ref{develop3} provides a balanced performance and is also \setlength{\fboxsep}{0pt}\colorbox{black!15}{grey-shaded}.

We plot $\text{NA}_{\text{seg}}$ vs. PESQ$(\tilde{\mathbf{s}})$ for Table\,\ref{develop3} as shown in Fig.\,\ref{fig:plot3}. It can be seen that our selected hyperparameter combination (solid blue line, asterisk markers) offers mostly very good (among the two best) PESQ$(\tilde{\mathbf{s}})$ and a strong noise attenuation, yielding a balanced performance.
\label{sec:5_1}
\subsection{Experimental Results and Discussion}
We report the experimental results on the test data for {\it seen} noises {\it types} (PED and CAFE), and {\it unseen} BUS noise separately. We investigate a CNN trained with the baseline losses, which are the conventional MSE, PW-FILT, and PW-PESQ, and with the newly proposed 2CL and 3CL losses. The measures on the {\it seen} noise types are shown in Tables\,\ref{PED_cafe_all} (all SNRs averaged) and \ref{PED_cafe_all_low} ($-5$\,dB SNR), the results on {\it unseen} BUS noise are shown in Tables\,\ref{BUS_all} (all SNRs averaged) and \ref{BUS_low} ($-5$\,dB SNR). The performance is averaged over all test speakers and if applicable all SNR conditions. In each column, the scheme offering the best performance is in {\bf boldface}. For the CNN trained with 2CL and 3CL, the selected settings are \setlength{\fboxsep}{0pt}\colorbox{black!15}{grey-shaded}, as shown in Tables\,\ref{develop} and \ref{develop2}, respectively.
\subsubsection{Seen Noise Types}
First, we look at the performance on the {\it seen} noise types as shown in Table\,\ref{PED_cafe_all}. It becomes obvious that the CNN trained by our proposed 2CL and 3CL offers mostly better SNR improvement than the CNN trained by the other baseline losses, reflected by a higher $\Delta$SNR. Among the CLs, 3CL offers the highest $\Delta$SNR on average. This is supposed to be attributed to the second term of both 2CL \eqref{CLF} and 3CL \eqref{CLF2} weighted by $\alpha$, representing the {\it filtered} noise component power, which is explicitly forced to be low during the training process. The CNN trained by PW-FILT loss also offers quite good noise attenuation, but with a poor residual noise quality, which is reflected by a very high WLAKR score. The proposed 3CL offers a very good, for CAFE also the best residual noise quality, as well as the strongest noise attenuation at the same time. This is expected, and is likely from the contribution of the third term in 3CL \eqref{CLF2}, which is supposed to preserve residual noise quality. During training, this term is explicitly forced to be low to keep a naturally sounding residual noise, by enforcing a similarity of the residual noise and the original noise component. Among the baseline methods, the CNN trained by PW-PESQ always shows the best residual noise quality. Surprisingly, the proposed 2CL also offers a better residual noise quality compared to the CNN trained with conventional MSE, even though the residual noise quality is not considered in the 2CL definition \eqref{CLF}. 

As introduced before, the CNN trained with conventional MSE tends to attenuate regions with very low SNR to optimize the global MSE \cite{shivakumar2016perception}, which may lead to strong noise distortion and speech component distortion. The proposed 2CL penalizes this speech component distortion by the first term of \eqref{CLF}, weighted by $1\!-\!\alpha$, which is not only good for preserving the speech component quality, but also for maintaining a naturally sounding residual noise. The CNNs trained by our proposed 2CL and 3CL by far provide the best speech component quality, which is reflected by at least 0.5\,dB higher SSDR and about 0.1 higher PESQ$(\tilde{\mathbf{s}})$ on average. This is attributed to the first term of 2CL \eqref{CLF} and 3CL \eqref{CLF2}, which is the loss function for the {\it filtered} speech component, and is supposed to preserve detailed structures of the speech signal and punishes the attenuation of the speech component. Among the CNNs trained by the components losses, 2CL offers slightly better PESQ$(\tilde{\mathbf{s}})$ and about 0.1\,dB higher SSDR compared to 3CL. One possible reason is that the weight for speech distortion in 3CL \eqref{CLF2} represented by $1\!-\!0.1\!-\!0.8\!=\!0.1$ is less compared to the one in 2CL \eqref{CLF} represented by $1\!-\!0.5\!=\!0.5$. {\it Our proposed 2CL and 3CL losses provide the best overall enhanced speech quality, which is reflected by obtaining the highest PESQ$(\hat{\mathbf{s}})$ and POLQA$(\hat{\mathbf{s}})$ scores.} In addition to that, 2CL and 3CL obtain slightly better speech intelligibility reflected by 0.01 higher STOI score for {\it seen} noise types on average. Among the CL-based CNNs, 3CL is better by offering a stronger noise attenuation, a more natural residual noise, and the best enhanced speech quality, yielding a more balanced performance. 
\begin{table}[t!]
	\centering
	\begin{subtable}{1\linewidth}
		\caption{Performance for {\bf seen noise types} (PED and CAFE) on the {\bf test set}; \textbf{All SNRs} averaged. Best approaches from Tables\,\ref{develop} and \ref{develop2} are \setlength{\fboxsep}{0pt}\colorbox{black!15}{grey-shaded}; Best scheme is in {\bf boldface}.}
		\Huge
		\setlength\tabcolsep{2pt}
		\centering
		\resizebox{1\linewidth}{!}{
			\begin{tabular}{|m{0.9cm}<{\centering}|c|c||c|c||c|c||c|c|c|}
				\hline
				\multirow{2}{*}{\stz{\rotatebox{90}{Noise}}} & \multicolumn{2}{|c||}{\multirow{2}{*}{Method}}                                                                           & \multicolumn{2}{c||}{Noise Component} & \multicolumn{2}{c||}{\stz{Speech Comp.}} & \multicolumn{3}{c|}{Total}        \\ \cline{4-10} 
				& \multicolumn{2}{|c||}{}                                                                                                   & \stz{$\Delta$SNR}     & WLAKR             & SSDR              & PESQ$(\tilde{\mathbf{s}})$  & PESQ$(\hat{\mathbf{s}})$  & POLQA$(\hat{\mathbf{s}})$ & STOI            \\ \hline
				\multirow{5}{*}{\rotatebox{90}{PED}} & \multicolumn{2}{|c||}{\stz{Baseline MSE}}                                                                                              &        5.84      &        0.24       & 11.70  &       2.94        &2.42  &  1.92 &   0.70       \\
				\hhline{~---------}
				&   \multicolumn{2}{|c||}{\stz{Baseline PW-FILT}}                                                                                              &        6.37     &        0.45       & 11.52  &       2.87        &2.53  &    1.91 &  0.70       \\
				\hhline{~---------}
				&   \multicolumn{2}{|c||}{\stz{Baseline PW-PESQ}}                                                                                              &        5.81     &       {\bf 0.16}       & 11.84  &       2.96        &2.47  &    1.89 &  0.70       \\
				\hhline{~---------}
				& \multicolumn{2}{|c||}{\stz{\cellcolor{black!15}{2CL\,($\alpha=0.5$)}}}                                                                                                &        \cellcolor{black!15}{6.18}      &     \cellcolor{black!15}{0.22}          &  \cellcolor{black!15}{\bf 12.34} & \cellcolor{black!15}{\bf 3.04}   &\cellcolor{black!15}{\bf 2.67}  &    \cellcolor{black!15}{2.11} &\cellcolor{black!15}{\bf 0.71}         \\ \hhline{~---------}
				& \multicolumn{2}{|c||}{\stz{\cellcolor{black!15}{3CL\,($\alpha=0.1,\,\beta=0.8$)}}}                                                                                                &        \cellcolor{black!15}{\bf 7.05}      &     \cellcolor{black!15}{0.18}          &  \cellcolor{black!15}{12.21} & \cellcolor{black!15}{3.00}   &\cellcolor{black!15}{\bf 2.67}  &  \cellcolor{black!15}{\bf 2.13}     &  \cellcolor{black!15}{\bf 0.71}\\ \hhline{----------}
				\multirow{5}{*}{\rotatebox{90}{CAFE}} & \multicolumn{2}{|c||}{\stz{Baseline MSE}}                                                                                              &        5.76      &        0.26       & 11.44  &       2.87        &2.33  &      1.90 &    0.69       \\
				\hhline{~---------}
				&   \multicolumn{2}{|c||}{\stz{Baseline PW-FILT}}                                                                                              &        6.32     &        0.60       & 11.42  &       2.84        &2.45  &   1.93 &   0.69       \\
				\hhline{~---------}
				&   \multicolumn{2}{|c||}{\stz{Baseline PW-PESQ}}                                                                                              &        5.78     &        0.21       & 11.57  &       2.90        &2.35  &   1.90 &   0.69       \\
				\hhline{~---------}
				& \multicolumn{2}{|c||}{\stz{\cellcolor{black!15}{2CL\,($\alpha=0.5$)}}}                                                                                                &        \cellcolor{black!15}{7.22}      &     \cellcolor{black!15}{\bf 0.13}          &  \cellcolor{black!15}{\bf 12.20} & \cellcolor{black!15}{\bf 3.05}   &\cellcolor{black!15}{2.60}  &    \cellcolor{black!15}{2.12} &\cellcolor{black!15}{\bf 0.70}         \\ \hhline{~---------}	
				& \multicolumn{2}{|c||}{\stz{\cellcolor{black!15}{3CL\,($\alpha=0.1,\,\beta=0.8$)}}}                                                                                                &        \cellcolor{black!15}{\bf 7.30}      &     \cellcolor{black!15}{\bf 0.13}          &  \cellcolor{black!15}{12.10} & \cellcolor{black!15}{3.03}   &\cellcolor{black!15}{\bf 2.62}  &  \cellcolor{black!15}{\bf 2.17} &  \cellcolor{black!15}{\bf 0.70}\\ \hhline{----------}
			\end{tabular}
		}
		\label{PED_cafe_all}
	\end{subtable}
	\newline
	\vspace*{0.4 cm}
	\newline
	\begin{subtable}{1\linewidth}
		\caption{Performance for {\bf seen noise types} (PED and CAFE) on the {\bf test set}; {\bf SNR$\mathbf{=-5}$\,dB}. Best approaches from Tables\,\ref{develop} and \ref{develop2} are \setlength{\fboxsep}{0pt}\colorbox{black!15}{grey-shaded}; Best scheme is in {\bf boldface}.}
		\Huge
		\setlength\tabcolsep{2pt}
		\centering
		\resizebox{1\linewidth}{!}{
			\begin{tabular}{|m{0.9cm}<{\centering}|c|c||c|c||c|c||c|c|c|}
				\hline
				\multirow{2}{*}{\stz{\rotatebox{90}{Noise}}} & \multicolumn{2}{|c||}{\multirow{2}{*}{Method}}                                                                           & \multicolumn{2}{c||}{Noise Component} & \multicolumn{2}{c||}{\stz{Speech Comp.}} & \multicolumn{3}{c|}{Total}        \\ \cline{4-10} 
				& \multicolumn{2}{|c||}{}                                                                                                   & \stz{$\Delta$SNR}     & WLAKR             & SSDR              & PESQ$(\tilde{\mathbf{s}})$  & PESQ$(\hat{\mathbf{s}})$  & POLQA$(\hat{\mathbf{s}})$ & STOI            \\ \hline
				\multirow{5}{*}{\rotatebox{90}{PED}} & \multicolumn{2}{|c||}{\stz{Baseline MSE}}                                                                                              &        6.10      &        0.24       & 3.03  &       1.83        &1.44  &     1.07 & 0.49       \\
				\hhline{~---------}
				&   \multicolumn{2}{|c||}{\stz{Baseline PW-FILT}}                                                                                              &        8.17      &        0.30       & 2.73  &       1.76        & 1.46  &      1.14 & 0.50       \\
				\hhline{~---------}
				&   \multicolumn{2}{|c||}{\stz{Baseline PW-PESQ}}                                                                                              &        6.43      &        {\bf 0.11}       & 3.00  &       1.85        & 1.45  &      {\bf 1.35} & {\bf 0.51}       \\
				\hhline{~---------}
				& \multicolumn{2}{|c||}{\stz{\cellcolor{black!15}{2CL\,($\alpha=0.5$)}}}                                                                                                &        \cellcolor{black!15}{8.25}      &     \cellcolor{black!15}{0.21}          &  \cellcolor{black!15}{\bf 3.10} & \cellcolor{black!15}{\bf 2.09}   &\cellcolor{black!15}{\bf 1.58}  &    \cellcolor{black!15}{1.28}& \cellcolor{black!15}{0.50}         \\ \hhline{~---------}
				& \multicolumn{2}{|c||}{\stz{\cellcolor{black!15}{3CL\,($\alpha=0.1,\,\beta=0.8$)}}}                                                                                                &        \cellcolor{black!15}{\bf 8.58}      &     \cellcolor{black!15}{0.21}          &  \cellcolor{black!15}{2.97} & \cellcolor{black!15}{2.05}   &\cellcolor{black!15}{\bf 1.58}  &  \cellcolor{black!15}{1.23}&  \cellcolor{black!15}{0.50}         \\ \hhline{----------}
				\multirow{5}{*}{\rotatebox{90}{CAFE}} & \multicolumn{2}{|c||}{\stz{Baseline MSE}}                                                                                              &        7.56      &        0.13       & 2.74  &       1.76        &1.39  &      {\bf 1.22} & 0.49       \\
				\hhline{~---------}
				&   \multicolumn{2}{|c||}{\stz{Baseline PW-FILT}}                                                                                              &        8.99      &        0.41       & 2.46  &       1.71        &1.42  &   1.07 &   0.49       \\
				\hhline{~---------}
				&   \multicolumn{2}{|c||}{\stz{Baseline PW-PESQ}}                                                                                              &       7.74      &        0.12       & 2.75  &       1.81        &1.42  &   {\bf 1.22} &   {\bf 0.51}       \\
				\hhline{~---------}
				& \multicolumn{2}{|c||}{\stz{\cellcolor{black!15}{2CL\,($\alpha=0.5$)}}}                                                                                                &        \cellcolor{black!15}{\bf 10.15}      &     \cellcolor{black!15}{0.11}          &  \cellcolor{black!15}{\bf 2.88} & \cellcolor{black!15}{2.04}   &\cellcolor{black!15}{\bf 1.57}  &    \cellcolor{black!15}{1.16} & \cellcolor{black!15}{0.50}         \\ \hhline{~---------}
				& \multicolumn{2}{|c||}{\stz{\cellcolor{black!15}{3CL\,($\alpha=0.1,\,\beta=0.8$)}}}                                                                                                &        \cellcolor{black!15}{9.93}      &     \cellcolor{black!15}{\bf 0.10}          &  \cellcolor{black!15}{2.84} & \cellcolor{black!15}{\bf 2.06}   &\cellcolor{black!15}{1.54}  &    \cellcolor{black!15}{1.17}& \cellcolor{black!15}{0.50}         \\ \hhline{----------}
			\end{tabular}
		}
		\label{PED_cafe_all_low}
	\end{subtable}	
\end{table}

The performance on the {\it seen} noise types at SNR$=-5$\,dB is shown in Table\,\ref{PED_cafe_all_low}. Both our proposed CLs and the baseline PW-FILT loss offer good noise attenuation, but the proposed CLs perform better. For CAFE noise, the proposed 2CL shows higher $\Delta$SNR compared to 3CL. Again, the baseline PW-FILT loss shows the worst residual noise quality reflected by very high WLAKR scores. Same as in Table\,\ref{PED_cafe_all}, the proposed 3CL and the baseline PW-PESQ provide the best residual noise quality for CAFE and PED noise, respectively. The proposed 2CL and 3CL offer the best speech component quality $(\text{PESQ}(\tilde{\mathbf{s}}))$ and overall enhanced speech quality $(\text{PESQ}(\hat{\mathbf{s}}))$. At SNR$=-5$\,dB, the CNN trained by the PW-PESQ loss offers slightly better speech intelligibility reflected by 0.01 higher STOI score compared to the CNNs trained by other losses. For {\it seen} noise types, our proposed 2CL and 3CL also provide the best enhanced speech quality in very harsh SNR conditions reflected by at least 0.1 points higher PESQ$(\hat{\mathbf{s}})$.
\subsubsection{Unseen Noise}
\begin{table}[t!]
	\centering
	\begin{subtable}{1\linewidth}
		\caption{Performance for {\bf unseen noise} (BUS) on the {\bf test set}; \textbf{All SNRs} averaged. Best approaches from Tables\,\ref{develop} and \ref{develop2} are \setlength{\fboxsep}{0pt}\colorbox{black!15}{grey-shaded}; Best scheme is in {\bf boldface}.}
		\Huge
		\setlength\tabcolsep{2pt}
		\centering
		\resizebox{1\linewidth}{!}{
			\begin{tabular}{|m{0.9cm}<{\centering}|c|c||c|c||c|c||c|c|c|}
				\hline
				\multirow{2}{*}{\stz{\rotatebox{90}{Noise}}} & \multicolumn{2}{|c||}{\multirow{2}{*}{Method}}                                                                           & \multicolumn{2}{c||}{Noise Component} & \multicolumn{2}{c||}{\stz{Speech Comp.}} & \multicolumn{3}{c|}{Total}        \\ \cline{4-10} 
				& \multicolumn{2}{|c||}{}                                                                                                   & \stz{$\Delta$SNR}     & WLAKR             & SSDR              & PESQ$(\tilde{\mathbf{s}})$  & PESQ$(\hat{\mathbf{s}})$ & POLQA$(\hat{\mathbf{s}})$ & STOI            \\ \hline
				\multirow{5}{*}{\rotatebox{90}{BUS}} & \multicolumn{2}{|c||}{\stz{Baseline MSE}}                                                                                              &        4.50      &        0.20       & 13.14  &       3.03        &2.39  &   2.39  & 0.71       \\
				\hhline{~---------}
				&   \multicolumn{2}{|c||}{\stz{Baseline PW-FILT}}                                                                                              &        5.56      &        0.39       & 13.43  &       3.12        &2.50 &      2.35  &{\bf 0.75}       \\
				\hhline{~---------}
				&   \multicolumn{2}{|c||}{\stz{Baseline PW-PESQ}}                                                                                              &        4.73      &        0.19       & 13.27  &       3.00        &2.42 &      2.34  &{\bf 0.75}       \\
				\hhline{~---------}
				& \multicolumn{2}{|c||}{\stz{\cellcolor{black!15}{2CL\,($\alpha=0.5$)}}}                                                                                                &        \cellcolor{black!15}{5.60}      &     \cellcolor{black!15}{\bf 0.18}          &  \cellcolor{black!15}{\bf 14.30} & \cellcolor{black!15}{\bf 3.38}   &\cellcolor{black!15}{2.63}  &    \cellcolor{black!15}{\bf 2.64}  &\cellcolor{black!15}{0.74}         \\ \hhline{~---------}
				& \multicolumn{2}{|c||}{\stz{\cellcolor{black!15}{3CL\,($\alpha=0.1,\,\beta=0.8$)}}}                                                                                                &        \cellcolor{black!15}{\bf 6.28}      &     \cellcolor{black!15}{\bf 0.18}          &  \cellcolor{black!15}{14.23} & \cellcolor{black!15}{3.35}   &\cellcolor{black!15}{\bf 2.68}  &   \cellcolor{black!15}{\bf 2.64}  & \cellcolor{black!15}{\bf 0.75}         \\ \hhline{----------}
			\end{tabular}
		}
		\label{BUS_all}
	\end{subtable}
	\newline
	\vspace*{0.4 cm}
	\newline
	\begin{subtable}{1\linewidth}
		\caption{Performance for {\bf unseen noise} (BUS) on the {\bf test set}; {\bf SNR$\mathbf{=-5}$\,dB}. Best approaches from Tables\,\ref{develop} and \ref{develop2} are \setlength{\fboxsep}{0pt}\colorbox{black!15}{grey-shaded}; Best scheme is in {\bf boldface}.}
		\Huge
		\setlength\tabcolsep{2pt}
		\centering
		\resizebox{1\linewidth}{!}{
			\begin{tabular}{|m{0.9cm}<{\centering}|c|c||c|c||c|c||c|c|c|}
				\hline
				\multirow{2}{*}{\stz{\rotatebox{90}{Noise}}} & \multicolumn{2}{|c||}{\multirow{2}{*}{Method}}                                                                           & \multicolumn{2}{c||}{Noise Component} & \multicolumn{2}{c||}{\stz{Speech Comp.}} & \multicolumn{3}{c|}{Total}        \\ \cline{4-10} 
				& \multicolumn{2}{|c||}{}                                                                                                   & \stz{$\Delta$SNR}     & WLAKR             & SSDR              & PESQ$(\tilde{\mathbf{s}})$  & PESQ$(\hat{\mathbf{s}})$ & POLQA$(\hat{\mathbf{s}})$ & STOI            \\ \hline
				\multirow{5}{*}{\rotatebox{90}{BUS}} & \multicolumn{2}{|c||}{\stz{Baseline MSE}}                                                                                              &        6.99      &        0.22       & 4.92  &       2.11        &1.55  &     1.59  & 0.61       \\
				\hhline{~---------}
				&   \multicolumn{2}{|c||}{\stz{Baseline PW-FILT}}                                                                                              &        8.03      &        0.24       & 5.07  &       2.17        &1.58 &    1.40  &  {\bf 0.64}       \\
				\hhline{~---------}
				&   \multicolumn{2}{|c||}{\stz{Baseline PW-PESQ}}                                                                                              &        7.05      &        {\bf 0.16}       & 4.86  &       2.05        &1.55 &    1.58  &  {\bf 0.64}       \\
				\hhline{~---------}
				& \multicolumn{2}{|c||}{\stz{\cellcolor{black!15}{2CL\,($\alpha=0.5$)}}}                                                                                                &        \cellcolor{black!15}{8.31}      &     \cellcolor{black!15}{0.20}          &  \cellcolor{black!15}{5.66} & \cellcolor{black!15}{2.57}   &\cellcolor{black!15}{1.73}  &    \cellcolor{black!15}{\bf 1.63}  &\cellcolor{black!15}{0.63}         \\ \hhline{~---------}
				& \multicolumn{2}{|c||}{\stz{\cellcolor{black!15}{3CL\,($\alpha=0.1,\,\beta=0.8$)}}}                                                                                                &        \cellcolor{black!15}{\bf 8.56}      &     \cellcolor{black!15}{0.21}          &  \cellcolor{black!15}{\bf 5.67} & \cellcolor{black!15}{\bf 2.60}   &\cellcolor{black!15}{\bf 1.77}  &   \cellcolor{black!15}{1.61}  & \cellcolor{black!15}{0.63}         \\ \hhline{----------}
			\end{tabular}
		}
		\label{BUS_low}
	\end{subtable}
	
\end{table}
The performance on the {\it unseen} BUS noise is shown in Tables\,\ref{BUS_all} and \ref{BUS_low}. Same as before, 2CL and 3CL provide very good noise attenuation and residual noise quality. The CNN trained by 3CL offers the highest $\Delta$SNR compared to the ones trained by other losses. Again, among the baselines, the PW-FILT loss always provides the highest $\Delta$SNR and the worst residual noise quality (high WLAKR). The PW-PESQ loss offers very good residual noise quality, sometimes even ranking best. Again, the proposed CLs clearly offer the best speech component quality $(\text{SSDR, PESQ}(\tilde{\mathbf{s}}))$ and total enhanced speech quality $(\text{PESQ}(\hat{\mathbf{s}}), \text{POLQA}(\hat{\mathbf{s}}))$. Especially, the proposed 3CL provides obviously better overall enhanced speech quality reflected by an about 0.2 points higher PESQ$(\hat{\mathbf{s}})$ compared to the other baseline losses. Except for the baseline MSE, the remaining baseline losses and our proposed CLs provide very comparable speech intelligibility as shown in the last column of Tables\,\ref{BUS_all} and \ref{BUS_low}. As before, the proposed 3CL performs best by offering good and balanced results.

In total, the CNN trained by our proposed components loss offers the best speech component quality for both {\it seen} and {\it unseen} noise types, in both averaged and very harsh noise conditions. At the same time, the two proposed CLs also offer the best $\Delta$SNR as well as a very good, in some cases even the best residual noise quality. So the CNN trained by our CLs show both a strong and a balanced performance by not only providing a strong noise attenuation, but also providing a naturally sounding residual noise, and a less distorted speech component. Likely from the contribution of all these aspects, our proposed CLs also provide the best enhanced speech quality and speech intelligibility in almost all experiments. Surprisingly, compared to the 2CL results, the additional third term in 3CL \eqref{CLF2}, which is supposed to preserve good residual noise quality, not only provides the same, sometimes even a better residual quality, but also indirectly increases noise attenuation during training. In total, the CNN trained by our 3CL offers the best and the most balanced performance.
\label{sec:5_2}
\section{Conclusion}
In this paper we illustrated the benefits of a components loss (CL) for mask-based speech enhancement. We introduced the 2-components loss (2CL), which controls speech component distortion and noise suppression separately, and also the 3-components loss (3CL), which includes an additional term to control the residual noise quality. Our proposed 2CL and 3CL are naturally differentiable for gradient-based learning, and do not need any additional training material or extensive computational effort compared to other auditory-related loss functions. Furthermore, we point out that these new loss functions are not limited to any specific network topology or application. In the context of a speech enhancement framework that uses a convolutional neural network (CNN) to estimate a spectral mask, the 3CL shows improvement over the baseline loss functions including the conventional MSE, the perceptual weighting filter loss, and the PESQ loss. On average, an at least 0.1 points higher PESQ score on the enhanced speech is obtained while also obtaining a higher SNR improvement by more than 0.5\,dB, for {\it seen} noise types. This improvement is even stronger for {\it unseen} noise types, where an about 0.2 points higher PESQ score is obtained on the enhanced speech while also the output SNR is ahead by more than 0.5\,dB. The new 2CL and 3CL loss functions are easy to implement and example code is provided at \href{https://github.com/ifnspaml/Components-Loss}{https://github.com/ifnspaml/Components-Loss}.
\label{sec:6}
\appendices
\section{}
{\bf Baseline PW-FILT}: The perceptual weighting filter applied in this loss function is borrowed from CELP speech coding, e.g., the adaptive multi-rate (AMR) codec~\cite{AMR3GPP}, in order to shape the coding noise\,/\,quantization error to be less audible by the human ear. This weighting filter is calculated according to~\cite{AMR3GPP} as
\begin{equation} \label{equ_wgh_filt}
W_{\ell}(z)=\frac{1-A_{\ell}(z/\gamma_1)}{1-A_{\ell}(z/\gamma_2)},
\end{equation}
with the predictor polynomial $A_{\ell}(z/\gamma)\!=\!\sum_{i=1}^{N_{\rm p}}\!a_{\ell}(i)\gamma^iz^{-i}$, $a_{\ell}(i)$ are the linear prediction (LP) coefficients of frame $\ell$, $N_{\rm p}$ is the prediction order, and $\gamma_1$, $\gamma_2$ are the perceptual weighting factors. During the search of the codebooks in CELP encoding, the error between the clean speech and the coded speech is weighted by the weighting filter and subsequently minimized. As a result, the weighted error becomes spectrally white, meaning that the final (unweighted) quantization error has a frequency distribution that is proportional to the frequency characteristics of the \textit{inverse} weighting filter $1/W_{\ell}(z)$, which has similarities to the shape of the clean speech spectral envelope. This property of the weighting filter allows to exploit the masking effect of the human ear: More energy of the quantization error will be placed in the speech formant region, where $1/W_{\ell}(z)$ is at some level below the spectral envelope \cite{zhao2019perceptual}.

After the original CELP weighting filter has been revisited, the corresponding perceptual weighting filter loss is now straightforward, which can be expressed as  
\begin{equation} \label{PW_loss}
J_{\ell}^{\text{PW-FILT}}=\sum_{k\in\mathcal{K}}\left | W_{\ell}(k) \right |^2\cdot\left ( \left | \hat{S}_{\ell}(k) \right | -\left | S_{\ell}(k) \right |\right )^2,
\end{equation}
where both $\hat{S}_{\ell}$ and $S_{\ell}(k)$ are effectively weighted by the weighting filter frequency response
\begin{equation} \label{weighting}
W_{\ell}(k)=W_{\ell}(z)\!\Bigm|_{z=e^{j\!\frac{2\pi k}{K}}}. 
\end{equation}
Similar to the original application of the weighting filter in speech coding, where the quantization error becomes less audible, the residual noise is also expected to be less audible compared to using the MSE loss. As a result, improved perceptual quality of the enhanced speech has been reported in \cite{zhao2019perceptual}.

{\bf Baseline PW-PESQ}: As with the standard PESQ, the PESQ loss as proposed in \cite{martin2018deep} consists of a symmetrical and an asymmetrical distortion, both are computed frame-by-frame in the loudness spectrum domain, which is closer to human perception \cite{martin2018deep}. The authors of \cite{martin2018deep} adopt the transformation operations from the amplitude spectrum domain to the loudness spectrum domain for the target and enhanced speech signals directly from the PESQ standard \cite{ITU862}. The symmetrical distortion $L_\ell^{(s)}$ for frame $\ell$ is obtained directly from the difference between the target and enhanced speech loudness spectra. Auditory masking effects should also be considered in calculating $L_\ell^{(s)}$. The corresponding asymmetrical distortion $L_\ell^{(a)}$ is computed based on the symmetrical distortion $L_\ell^{(s)}$, but weighting the positive and negative loudness differences differently. Because human perceptions of the positive and negative loudness differences are not the same. Thus, different auditory masking effects must be considered, respectively. Then the PESQ loss is defined as 
\begin{equation} \label{pesq_loss}
J_{\ell}^{\text{PESQ}}=\theta_1\cdot L_\ell^{(s)}+\theta_2\cdot L_\ell^{(a)},
\end{equation}
where $\theta_1$ and $\theta_2$ are weighting factors, and are set to $0.1$ and $0.0309$, respectively \cite{martin2018deep}. Since $J_{\ell}^{\text{PESQ}}$ is highly non-linear and not fully differentiable, the authors propose to combine the PESQ loss with the conventional MSE that is fully differentiable as the final loss to make the gradient-based learning more stable. Thus, the used loss function for training is defined as \cite{martin2018deep}
\begin{equation} \label{PWPESQ_loss}
J_{\ell}^{\text{PW-PESQ}}=\lambda_1\cdot J_{\ell}^{\text{MSE}} +\lambda_2\cdot J_{\ell}^{\text{PESQ}},
\end{equation}
with $J_{\ell}^{\text{MSE}}$ directly calculated from \eqref{loss}, $\lambda_1\in\left [ 0,1 \right ]$ and $\lambda_2\in\left [ 0,1 \right ]$ being the weighting factors for the MSE loss and the PESQ loss, respectively. The network trained by this loss function not only aims at a low MSE loss, but also needs to decrease the distortions.

{\bf Baseline PW-STOI}: To mimic the frequency selectivity of the human ear, the amplitude spectra of the target clean speech and the predicted speech need to be framewise transformed to one-third octave bands, as proposed in \cite{kolbcek2018monaural}. Therefore, let $S_\ell^{\text{oct}}(b)$ and $\hat{S}^{\text{oct}}_\ell(b)$ be the one-third octave band decompositions for the clean speech and the enhanced speech, respectively, with $\ell$ being the frame index and the one-third octave band index $b\in\mathcal{B}=\left \{ 0,1,\ldots,B\!-\!1 \right \}$. The $b\text{th}$ decomposition of the target clean speech can be obtained by
\begin{equation} \label{STOI_octave}
S_\ell^{\text{oct}}(b)=\sqrt{\sum_{k\in \mathcal{K}_b}\left | S_\ell(k) \right |^2},
\end{equation}
where $\mathcal{K}_b$ denotes the set of DFT frequency bin indices of the $b\text{th}$ one-third octave band, which are specified in \cite{kolbcek2018monaural}. Similarly, $\hat{S}_\ell^{\text{oct}}(b)$ is obtained by the same operation as \eqref{STOI_octave}, just replacing $S_\ell(k)$ by the predicted one $\hat{S}_\ell(k)$. A number of $B=15$ one-third octave bands are used \cite{kolbcek2018monaural}. Then, a target speech short-time temporal envelope vector
\begin{equation} \label{STOI_octave_vector}
\mathbf{S}_\ell^{\text{oct}}(b)=\left[S_{\ell-N+1}^{\text{oct}}(b), S_{\ell-N+2}^{\text{oct}}(b), \ldots, S_{\ell}^{\text{oct}}(b) \right]^\intercal
\end{equation}
for the $b\text{th}$ one-third octave band is generated, with $N=30$ to capture the important modulation frequencies \cite{kolbcek2018monaural}, and $\left[\cdot\right]^\intercal$ being the transpose. The vector $\mathbf{\hat{S}}_\ell^{\text{oct}}(b)$ is obtained analogously. The differentiable STOI approximation for the $b\text{th}$ band is finally defined as
\begin{equation} \label{STOI_approx}
L_{\ell}^{\text{STOI}}(b)=\frac{\left ( \mathbf{S}_\ell^{\text{oct}}(b)- \bm{\mu}_\ell^{\text{oct}}(b) \right )^\intercal \cdot \left ( \mathbf{\hat{S}}_\ell^{\text{oct}}(b)- \bm{\hat{\mu}}_\ell^{\text{oct}}(b) \right )}{\left \|  \mathbf{S}_\ell^{\text{oct}}(b)- \bm{\mu}_\ell^{\text{oct}}(b) \right \| \cdot \left \| \mathbf{\hat{S}}_\ell^{\text{oct}}(b)- \bm{\hat{\mu}}_\ell^{\text{oct}}(b) \right \|},
\end{equation}
with $\left \| \cdot \right \|$ being the $\ell^2$-norm operation, $\cdot$ being the dot product, and $\bm{\mu}_\ell^{\text{oct}}(b)$ and $\bm{\hat{\mu}}_\ell^{\text{oct}}(b)$ being the sample means of the vectors $\mathbf{S}_\ell^{\text{oct}}(b)$ and $\mathbf{\hat{S}}_\ell^{\text{oct}}(b)$, respectively.

During training, the network should maximize this STOI approximation. So the STOI loss is defined as the negative of $L_{\ell}^{\text{STOI}}(j)$ which needs to be minimized in the training phase:
\begin{equation} \label{PWSTOI_loss}
J_{\ell}^{\text{PW-STOI}}=-\frac{1}{B} \cdot{\sum_{b\in \mathcal{B}}L_{\ell}^{\text{STOI}}(b)}.
\end{equation}

\ifCLASSOPTIONcaptionsoff
  \newpage
\fi

\bibliographystyle{IEEEtran}
\bibliography{main}
%

%
%
%




\end{document}